\documentclass[twocolumn]{openjournal} 

\usepackage{orcidlink} 
\usepackage[T1]{fontenc}
\usepackage{ae,aecompl}
\usepackage[T1]{fontenc}
\usepackage{ae,aecompl}
\usepackage{hyperref}
\usepackage{url}

\usepackage{longtable}
\usepackage{booktabs}
\usepackage{tabu}
\usepackage{graphicx}
\usepackage{multirow}
\usepackage{ulem}
\usepackage{color}
\usepackage{amsmath}

\def \cm{~\rm{cm}}
\def \s{~\rm{s}}
\def \km{~\rm{km}}

\def \g{~\rm{g}}

\def \erg{~\rm{erg}}

\def \yr{~\rm{yr}}

\def \keV{~\rm{keV}}

\usepackage{xcolor}
\definecolor{redak}{rgb}{0.9,0.15,0.05}

\shorttitle{Simulating the JJEM: circum-jet rings}
\shortauthors{Akashi and Soker}

\graphicspath{{./}{figures/}}

\begin{document}

\title{Simulating the jittering-jets explosion mechanism: circum-jet rings account for observed core-collapse supernova remnant morphologies}

\author{Muhammad Akashi\,\orcidlink{0000-0001-7233-6871}}
\affiliation{Kinneret College on the Sea of Galilee, Samakh 15132, Israel}
\affiliation{Department of Physics, Technion - Israel Institute of Technology, Haifa, 3200003, Israel; akashi@technion.ac.il}

\author{Noam Soker\,\orcidlink{0000-0003-0375-8987}} 
\affiliation{Department of Physics, Technion - Israel Institute of Technology, Haifa, 3200003, Israel; soker@technion.ac.il; }


\begin{abstract}
We conduct three-dimensional hydrodynamical simulations of core-collapse supernova (CCSN) explosion driven by jets in the framework of the jittering jets explosion mechanism (JJEM), and obtain a pair of opposite circum-jet rings similar to those observed in some CCSN remnants (CCSNRs). We launch two pairs of jets along the same axis, the first of two opposite wide jets, and the second of narrow jets. The wide jets compress the core of a stripped-envelope stellar model to form a dense, fast-expanding shell. The narrow jets catch up with the dense shell, penetrate it, and compress the gas to the sides, forming the two opposite rings. At high inclination angles of the jets' axis to the line of sight, the projection of each ring on the plane of the sky forms two bright zones, where the rings cross the plane of the sky. This morphology explains that of SNR G46.8-0.3. At intermediate inclination angles, the rings are fully visible as two opposite bright elliptical rims. Our simulations explain the two prominent rings on the outer shell of CCSNR G11.2-0.3. Our results strengthen the claim that the JJEM is the primary explosion mechanism of CCSNe.                                              
\end{abstract}
   

\section{Introduction} 
\label{sec:intro}

In the jittering jets explosion mechanism (JJEM) of core-collapse supernovae (CCSNe), pairs of jets with fully or partially stochastic variations in jets' directions explode the massive star (for recent reviews of the JJEM properties and its outcomes, see \citealt{Soker2024UnivReview} and \citealt{Soker2025Learning}, and for its relation to other energy sources in the explosion see \citealt{Soker2025G11}). 
Stochastically intermittent accretion disks around the newly born neutron star (NS) launch several to $\approx 20$ pairs of jittering jets during the explosion. A partially stochastic behavior occurs when the rotation of the pre-collapse core is non-negligible, leading to jittering around the direction of the core's angular momentum axis. Vortices in the convection zones of the collapsing core are the source of the stochastic angular momentum fluctuation seeds that instabilities above NS amplify (e.g., \citealt{GilkisSoker2014, ShishkinSoker2023, WangShishkinSoker2024, WangShishkinSoker2025}; for studies of these types of instabilities see, e.g., \citealt{Abdikamalovetal2016, KazeroniAbdikamalov2020, Buelletetal2023}).

One of the strong supports of the JJEM comes from analysis of CCSN remnants (CCSNRs) that exhibit jet-shaped morphologies, particularly point-symmetric morphologies (e.g., \citealt{BearSoker2025, SokerShishkin2025Vela, ShishkinSoker2025Crab, Soker2025G0901, Soker2025G11}). Point-symmetrical morphologies include two or more pairs of opposite structural features that do not share a common axis through the center (e.g., \citealt{ShishkinMichaelis2026}). The structural features might be bubbles, rings, lobes, clumps, filaments, nozzles, and ears, as three-dimensional simulations of the JJEM show \citep{Braudoetal2025, Braudoetal2026, SokerAkashi2025, AkashiSoker2026, BraudoSoker2026}.  Hydrodynamical simulations of the neutrino-driven mechanism do not produce point-symmetric morphologies (e.g., \citealt{OrlandoJankaetal2025A, OrlandoJankaetal2025B, OrlandoMicelietal2025, Orlando2026}).  
In almost all CCSNe, according to the JJEM, the jittering jets are not relativistic (\citealt{Guettaetal2020} claim that most CCSNe do not have relativistic jets, unlike relativistic jets of gamma ray bursts, e.g., \citealt{Izzoetal2019, AbdikamalovBeniamini2025}).
Recent studies show that in some cases there are one, two, or three very powerful pairs of jets, and that one jet might be much more powerful than the opposite jet in a pair (e.g., \citealt{Bearetal2025Puppis, Shishkinetal2025S147}). Such two (or three) energetic pairs of jets can form two (or three) photospheric shells during a supernova explosion, as identified in SN 2023ixf by \cite{SokerShiran2025} based on observation by \cite{Zimmermanetal2024} and by \cite{ShiranSoker2026} in SN 2024ggi based on observations by \cite{ChenTWetal2025}. 

Generally, supernova remnants (SNRs) can reveal valuable information about the physics of supernovae and their interaction with the ambient gas, interstellar medium (ISM) and circumstellar material (CSM) (e.g., \citealt{Yanetal2020RAA, Luetal2021RAA, Soker2025UnivTycho, Kundu2026}), involving jets (e.g., \citealt{YuFang2018RAA}), ejecta's structure (e.g., \citealt{Renetal2018RAA}), emission properties and formation of cosmic rays (e.g., \citealt{Yamazakietal2014RAA, Zhangetal2016RAA, Lietal2020RAA, Cristofari2021Univ, Ranasingheetal2021Univ, Leahy2022Univ, Leahyetal2022Univ, SinitsynaSinitsyna2023Univ, Zhaoetal2023Univ, GiulianiCardillo2024Univ, Luoetal2024RAA, Aktekinetal2026}), magnetohydrodynamical effects (e.g., \citealt{Wuetal2019RAA, Leietal2024RAA}), and the roles of the NS remnant  (e.g., \cite{HorvathAllen2011RAA, Wuetal2021RAA, Olmi2023Univ, Popov2023Univ}). In recent years, it has become clear that CCSNRs may be the best objects to point at the explosion mechanism of CCSNe: whether it is the JJEM (e.g., \citealt{Soker2025Learning}) or the alternative neutrino-driven mechanism (e.g., \citealt{Burrowsetal2024, Janka2025}).

The above-described potential of CCSNRs to reveal the nature of the explosion process, and in particular to exhibit jet-shaped morphologies that hydrodynamical simulations can reproduce, motivated this study. We conducted a set of specific numerical simulations of two pairs of jets, one after the other, that share the same symmetry axes and differ by their opening angle.  We obtained, among others, specific morphological features that resemble structures in the CCSNRs G46.8-0.3 (also called HC 30; Section \ref{sec:G46}) and G11.2-0.3 (Section \ref{sec:Rings}). We describe our numerical setting in Section \ref{sec:Numerics}, and the relevant results of these simulations in Section \ref{sec:Results}. 
We note that \cite{GarciaSeguraetal2021} simulated alternative narrow and wide jets in post common envelope binary systems. Such jets can form a wide variety of bipolar shapes and morphological features (e.g., \citealt{GarciaSeguraetal2022}). We focus on the shaping of a pair of rings and their appearance at different inclination angles.  
We summarize this study in Section \ref{sec:Summary}. 

\section{The numerical settings}
\label{sec:Numerics}
\subsection{Launching two pairs of jets}
\label{subsec:TwoPairs}
We simulate the formation of bipolar circum-jet rings resulting from the interaction of jets with a model of the core of a massive star. In our previous work \citep{SokerAkashi2025}, we assumed that earlier jets exploded the core before the pair of late jets we launched, and we imposed a dense, fast-expanding shell in the core as an initial condition to the hydrodynamical simulations. Here, we launch two consecutive pairs of early conical jets into the core; these are the first jets in the explosion. Due to limited numerical resources, we made some simplifications. (1) Although the jet's origin is tens of km from the center, we launch the jets at thousands of km. (2) We neglect gravity. For that, we launch very energetic jets that accelerate the inner core in a short time to velocities much larger than the escape velocities in the interaction regions at thousands of km. (3) We do not inject the later jets expected in the JJEM. We will, therefore, not consider the material near the center of the ejecta that later jets should accelerate. We focus on the structures formed by the pair of jets in the ejecta's outskirts.  

We perform three-dimensional hydrodynamical simulations using the Eulerian adaptive-mesh refinement (AMR) code \textsc{FLASH} v4.8 \citep{FryxellEtAl2000}, employing the unsplit hydrodynamics solver. The computational domain is a Cartesian box $(x,y,z)$ with
\begin{equation}
    -1.8 \times 10^{10} \cm \le x,y,z \le 1.8 \times 10^{10} \cm,
\label{eq:size}
\end{equation}

corresponding to a cubic domain of side length
\begin{equation}
   L = 3.6 \times 10^{10} \cm.
 \label{eq:length}
\end{equation}

We apply outflow boundary conditions on the six cube faces.
We use 7 refinement levels above the base grid, with one additional enforced level in the central injection region (maximum of 8 levels), giving an effective resolution of $2^{10}$ cells per dimension and a minimum cell size of
\begin{equation}
    \Delta x_{\min} = 3.5 \times 10^{7}\ \mathrm{cm}.
 \label{eq:deltax}
\end{equation}
Refinement is based on density and pressure gradients.

\begin{figure}
\centering
\includegraphics[trim=0.0cm 7.3cm 0.0cm 7.5cm ,clip, scale=0.45]{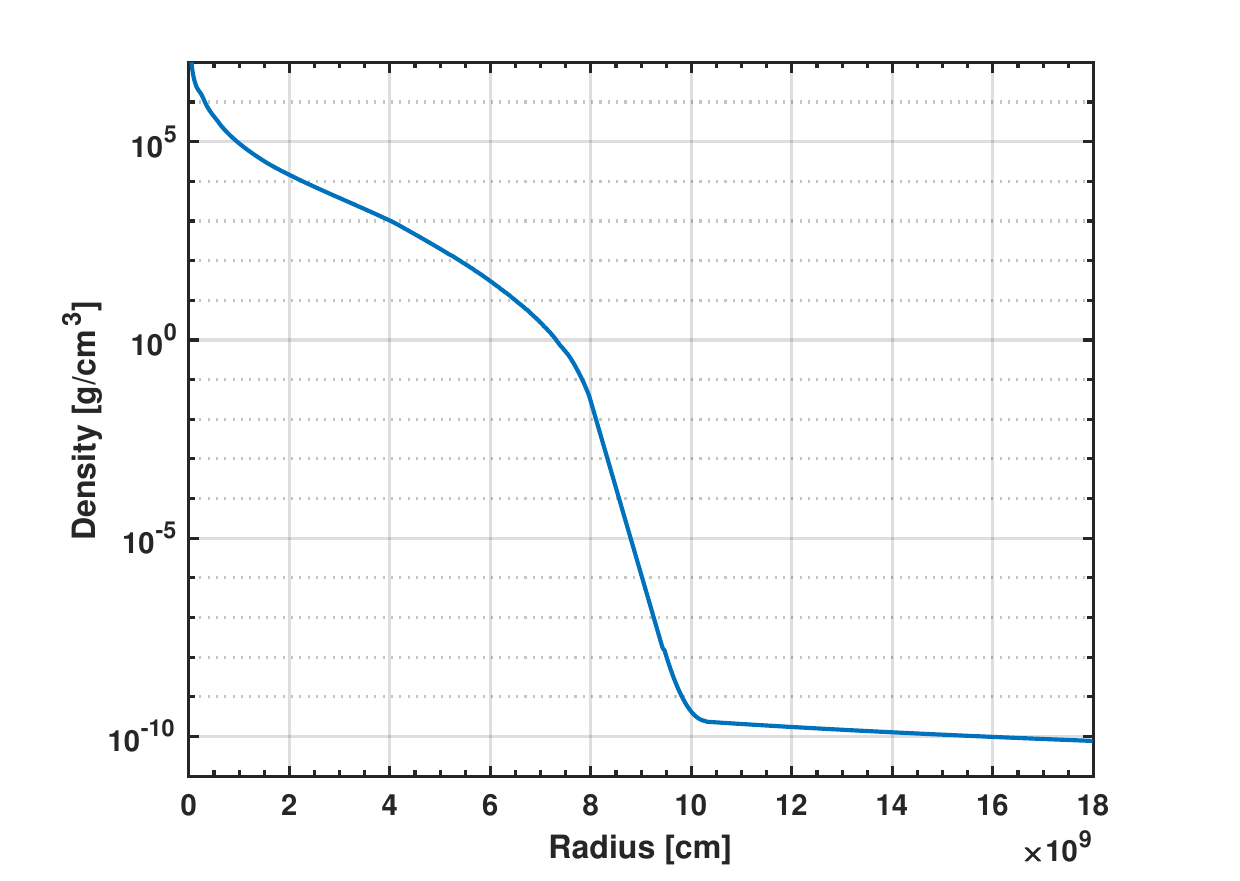} 
\caption{  The density profile at $t=0$. The density profile includes a smoothing of the stellar profile and the outer stellar wind profile around the stellar radius of $R_\ast= 8\times10^{9} \cm = 0.115 R_\odot$. The composition is uniform in the entire grid.  
}
\label{fig:DensityProfile}
\end{figure}

At $t=0$, we set at the central volume of the computation cube a spherical striped-envelope stellar model adapted from \cite{Braudoetal2025}. 
The innermost region of this stellar model is from \citet{PapishSoker2014Planar}, who fit a post-bounce structure of a $15 M_\odot$ progenitor at $t \simeq 0.2 \s$ after bounce \citep{Liebendorferetal2005}. This inner profile extends to radii of $ 2\times10^{9}\cm$. Between $2\times10^{9} \cm$ and the stellar surface at $8\times10^{9} \cm$, there is an envelope of a hydrogen- and helium-stripped $15 M_\odot$ Wolf–Rayet progenitor obtained from a MESA simulation. 
 The stellar radius is $R_\ast= 8\times10^{9} \cm = 0.115 R_\odot$. It is a hydrogen and helium-stripped star from a progenitor of $15 M_\odot$. Such a star requires a binary interaction to remove its envelope (e.g., \citealt{Gilkisetal2025}). We present the density profile in Figure \ref{fig:DensityProfile}. The density outside the star sharply drops to a lower value connected to a density that decreases as that of a stellar wind.  
To avoid numerical difficulties near the origin, we impose an inner inert core of radius
\begin{equation}
    R_{\rm core} = 4 \times 10^{8} \cm,
\label{eq:Rcore}
\end{equation}
within which we set the velocity to zero.
%

We initiate the explosion by injecting a pair of opposite wide-angle jets during the first half-second,
\begin{equation} 
\Delta t_{\rm wide} = 0.5 \s.
\label{eq:WideDt}
\end{equation}
Each jet of the first pair has a half-opening angle of 
\begin{equation} 
\alpha_{\rm wide,1} = 60^\circ
\label{eq:WideAngle}
\end{equation}
and a velocity of 
\begin{equation} 
v_{\rm wide} = 8 \times 10^{4} \km \s^{-1}, 
\label{eq:WideVelocity}
\end{equation}
The total kinetic energy of the two jets combined is  
\begin{equation} 
E_{\rm wide} = 3 \times 10^{51} \erg.
\label{eq:Wideenergy}
\end{equation}
This wide outflow drives strong shocks and inflates a dense expanding shell.

At a later time $t_2$, we launch a second pair of jets that interact with the structure that the first pair shaped. Given the very energetic explosion we simulate, much greater than the typical $\simeq 10^{51} \erg$, we also set a relatively long explosion time. We present most results for $t_2=3 \s$, longer than the typical explosion time of $\simeq 1 \s$. We also conduct simulations with  $t_2 = 1 \s$. 
We launch the second pair of jets with a velocity of 
\begin{equation} 
v_{\rm narrow} = 8 \times 10^{4} \km \s^{-1},
\label{eq:Vnarrow}
\end{equation}
injected within a radius
\begin{equation} 
r_{\rm inj} = 7 \times 10^{8} \cm.
\label{eq:Rinj}
\end{equation}
 This is the same injection radius as of the wide jets. During the jet-launching phase, for both narrow and wide jets, the hydrodynamic quantities inside the injection region, a cone on each side, within the prescribed opening angles of the jets, are overwritten ("reset") at every timestep for the duration of the injection. Specifically, the jet density and velocity are imposed directly in these regions to match the prescribed jet properties. The thermal energy is not modified during the injection and remains equal to that of the original ambient medium at the corresponding location. The thermal quantities are fully consistent with the equation of state throughout the simulations.

We conduct several simulations that differ in the properties of the second pair of jets, the narrow jets; we summarize them in Table \ref{Table1}. To save computation time in scanning the parameter space, the two opposite jets, the up and down jets, have different properties. To minimize the effect of the Cartesian grid, in all simulations the narrow-jet axis is inclined at $30^\circ$ to the $z$-axis, lying in the $xz$ plane.
The symmetry axis of the wide jets is the same in all simulations (marked A, for aligned), except for one that we mark M (for misaligned). In Run M, the symmetry axis of the first wide pair is along the $z$ axis.  
\begin{table}[t]
\begin{center}
\caption{Summary of the simulation runs.}
\begin{tabular}{lccccc}
\hline
Run & $t_2$ & $\alpha_{\rm u}$ & $\alpha_{\rm d}$ & $E_{\rm u}$ & $E_{\rm d}$  \\
    & s     &                  &                  &      erg    &   erg        \\ 
\hline
A  & 3 & $10^\circ$ & $5^\circ$ & $1\times10^{50}$ & $1\times10^{50}$ \\
A2 & 3 & $10^\circ$ & $5^\circ$ & $3.3\times10^{49}$ & $3\times10^{50}$ \\
A3 & 3 & $20^\circ$ & $20^\circ$ & $3.3\times10^{49}$ & $3\times10^{50}$ \\
AEi & 1 &   $10^\circ$ & $5^\circ$ & $1\times10^{50}$ & $1\times10^{50}$ \\
AEii& 2 &   $10^\circ$ & $5^\circ$ & $1\times10^{50}$ & $1\times10^{50}$ \\
M  & 3 & $10^\circ$  & $5^\circ$ & $1\times10^{50}$ & $1\times10^{50}$ \\
\hline
\label{Table1} 
\end{tabular}
\end{center}
\begin{flushleft}
\small 
Notes: In all simulations, the two opposite wide jets, the first pair, have the same properties as in equations (\ref{eq:WideDt}) - (\ref{eq:Wideenergy}). We start launching the first pair at $t=0$ until $t=0.5 \s$. The axis of the second pair of jets, the narrow jets, is the same in all simulations: in the $xz$ plane inclined at $30^\circ$ to the $z$-axis. In Runs A, A1, A3, AEi, and AEii, the axis of the wide jets is aligned with that of the narrow jets. Only in Run M are the axes misaligned by $30^\circ$ and, with the wide jets' axis along the $z$-axis. The launching velocity of all jets is $8 \times 10^4 \km \s^{-1}$, and the launching period of the narrow jets is $\Delta t_{\rm narrow} = 1 \s$.  
Variables: $t_2$ is the time when we start launching the second pair of jets. $\alpha_{\rm u}$, $\alpha_{\rm d}$, $E_{\rm u}$, and $E_{\rm d}$ are the half-opening angle and energy of the up and down jets, respectively.    
\end{flushleft}
\end{table}

\subsection{Justification}
\label{subsec:Justification}

 Several comments are in place here regarding our assumptions and settings.

 We launch a wide pair of jets to set the explosion. A wide jet can stand for several narrow jets jittering around an axis, or a jet precessing around an axis. It is also possible that an episode of mass accretion via an accretion disk will launch two opposite wide jets. The jet-launching episodes in the JJEM are short, and in many cases, the accretion disk might have no time to relax, leading to the launching of unequal jets in a pair, or not-well-collimated jets (e.g., \citealt{Soker2024CounterJet}). In those cases, the outflow can be wide to start with. The half-opening angle of a jet at launching might be $<60^\circ$. After breaking out of the stalled shock and interacting with the falling core material at distances of hundreds to a few thousand kilometers, it forms a wider cocoon. This is a process for a future simulation. We launch the jets at thousands of kilometers, so they are already launched as wide jets.

 There is another justification for a small number of energetic pairs of jets. Recent studies of the JJEM concluded that some CCSNR morphologies indicate shaping by two or three very energetic pairs of jets, e.g., SNR S147 for which  \cite{Shishkinetal2025S147} deduced an explosion energy of $1.5-2 \times 10^{51} \erg$, larger than the canonical explosion energy. In this study, we set explosion energies of $3.2-3.33 \times 10^{51} \erg$. Such energies are observed in some CCSNe, and so are not outside the observed range. For example, \cite{UtrobinChugai2026} inferred the explosion energy of SN 2023ixf to be $2.8 \times 10^{51} \erg$.

 The large explosion energy leads to velocities of $>3 \times 10^{4} \km \s^{-1}$, which is twice and more the escape velocity from $r=10,000 \km$. This justifies the neglect of gravity. Future simulations will be more accurate and include gravity. These will require longer simulation time. For the present exploratory study of ring formation, we sacrificed accuracy to allow scanning of the parameter space. In any case, we do not expect gravity to alter our conclusions regarding the formation of pairs of rings. We here focus on the formation of rings and do not study other aspects of the JJEM that require including gravity, such as the 3D simulations by \cite{Braudoetal2026} that launch jets closer to the center and so must include gravity.

 In principle, other types of models with well collimated jets might form rings, like Magnetorotational-driven CCSNe (e.g.,  
\citealt{ObergaulingeraLOY2021}), relativistic jets from a black hole (e.g., \citealt{Gottliebetal2022}) and magnetars (e.g., \citealt{ChenKJetal2020}). Simulations of these types of explosions should show that they can indeed form rings. However, we do not consider these mechanisms to be likely alternatives to the JJEM, because most CCSNRs with pairs of rings along an axis have another symmetry axis that is largely inclined to the rings' axis, like SNR G11.2-0.3 that we discuss in Section \ref{sec:Rings} and the SNR Circinus X-1 for which we simulated the rings in \cite{SokerAkashi2025}. The JJEM explains two or more highly inclined symmetry axes, whereas the aforementioned mechanisms predict a nearly fixed axis. According to the JJEM, there are more pairs of jets than the two we launch in our simulations. But here we launch only two to form the jets, and do not reproduce the structural feature along other axes (see \citealt{AkashiSoker2026} for another pair of jets).

\section{Results}
\label{sec:Results}

To reveal the outflow properties, we present some physical quantities in a plane through the center of the grid. For comparison with observations (Sections \ref{sec:G46} and \ref{sec:Rings}), we present the numerical emission integral  
\begin{equation}
EI= \int \rho^2 dl = \int \rho^2 dY_s,
\label{eq:EI}
\end{equation}
where $\rho$ is the density, and $dl$ is an element along the line of sight; here we take the $Y_s$ to be line-of-sight coordinate, $d Y_s = dl$, such that $X_s Z_s$ is the plane of the sky. 

We first describe the evolution of Run A and some of its physical properties. We then present the emission integral of the other simulations. 

\subsection{The flow structure of Run A}

\begin{figure}
\centering
\includegraphics[width=\linewidth, trim=4.5cm 0.0cm 4.5cm 0cm, clip]{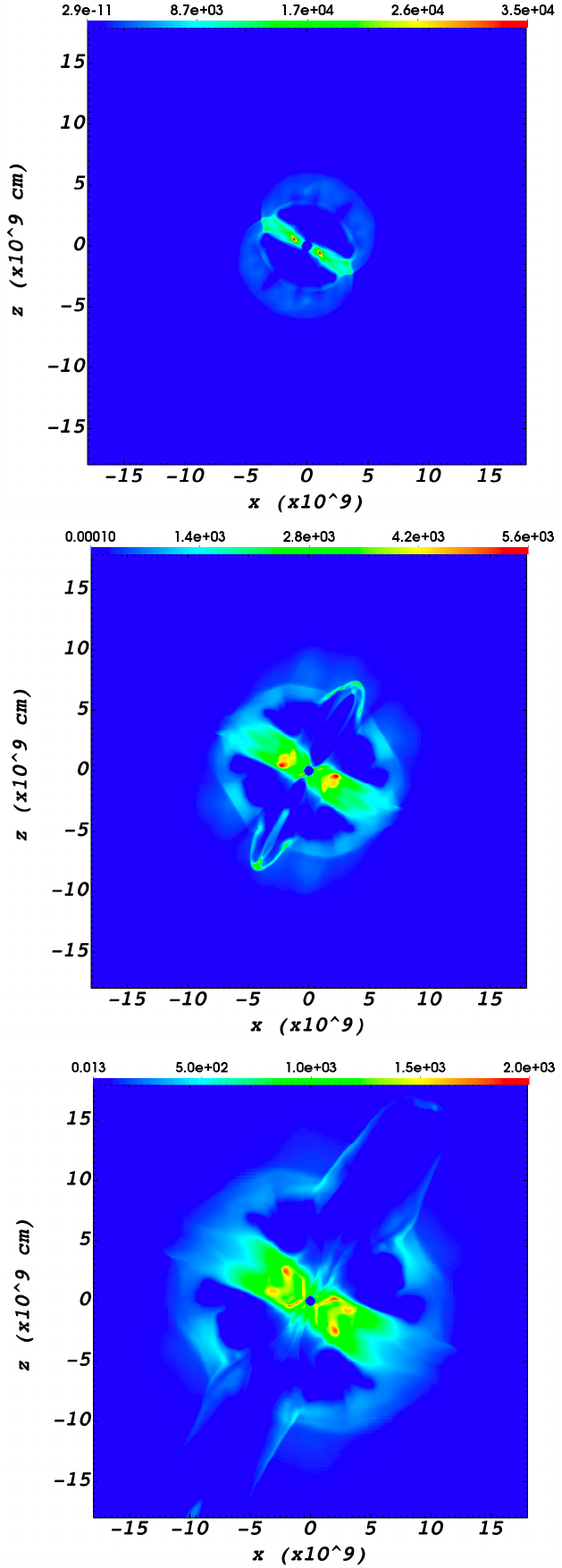}
\caption{ Time evolution of the density in the meridional $xz$ plane for Run A at $t=3$, $4.4$, and $6.4 \s$. The upper panel shows only the effect of the first pair of jets (the wide pair launched from $t=0$ to $t= 0.5 \s$), as we start launching the narrow jets at $t_2=3 \s$, until $t=4 \s$. The density is shown on a linear scale in $\g \cm^{-3}$.}
\label{fig:density_evolution}
\end{figure}

\begin{figure}
\centering
\includegraphics[width=\linewidth, trim=4.5cm 1.0cm 4.5cm 0cm, clip]{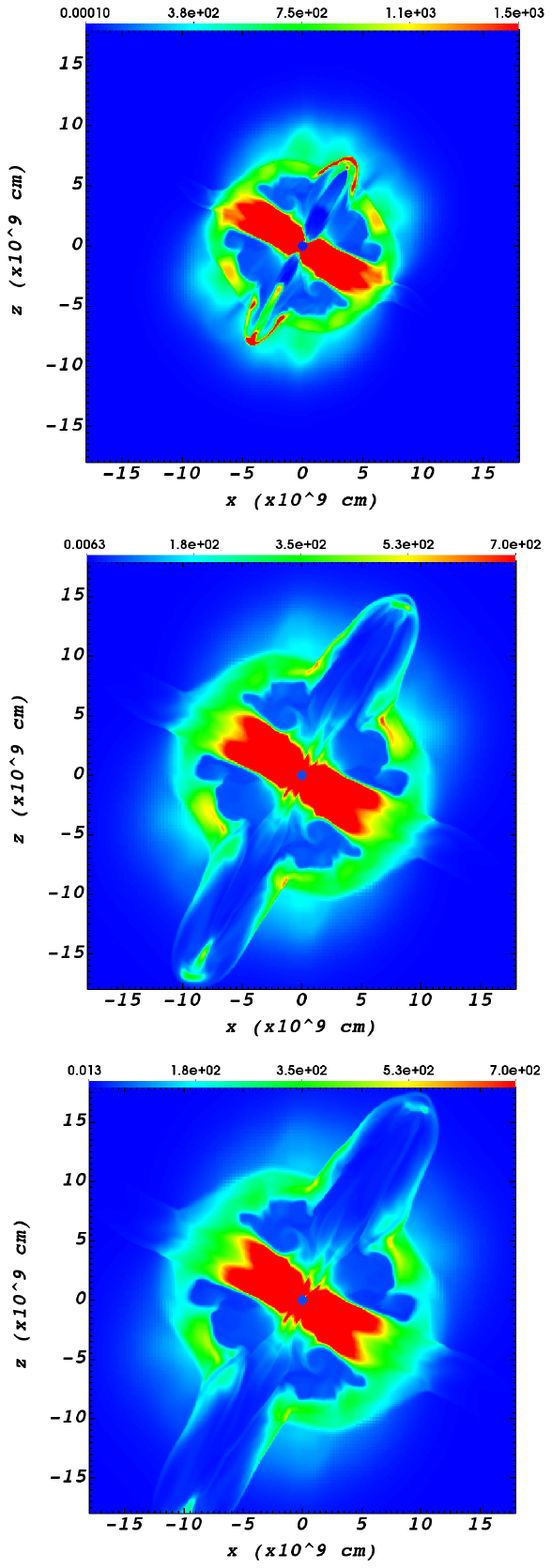}
\caption{Similar to Figure \ref{fig:density_evolution}, but on a different density color-scale and at times of $t=4.4$, $6$, and $6.4 \s$. The density scaling here emphasizes the outer regions that we study here (we do not study the dense inner region in red).  There are two low-density cavities (blue regions): to the upper right and lower left of the red dense region, at the base of the two long lobes. The boundaries of cavities are not smooth; rather, there are several `tongues' (in green). These result from Rayleigh-Taylor instability, as the cavity surroundings are unstable, as we discuss later. The low resolution suppresses short-wavelength instability modes.  
}
\label{fig:density_evolution_2}
\end{figure}

Figure~\ref{fig:density_evolution} shows the time evolution of the density in the $xz$ plane for Run A at $t=3 \s$, just as we start launching the second pair of jets,  $t=4.4 \s$, and $t=6.4 \s$. The density maps are on a linear scale in units of $\g \cm^{-3}$. The symmetry axis of the pairs of jets is in this plane at $30^\circ$ to the $Z$ axis. 
As we discussed in Section \ref{sec:Numerics}, due to numerical limitations, we do not simulate more jets, which, according to the JJEM, are supposed to further accelerate the material that stayed near the center after the first two pairs of jets. Therefore, in this study, we examine only the morphology of the outer ejecta zone, namely, the dense shell and the outer region.
For that, in Figure \ref{fig:density_evolution_2} we use a different scale to emphasize the outer regions; the red-colored high-density zone near the center is not studied here. Note that the three times of Figure \ref{fig:density_evolution_2} are not exactly as in Figure \ref{fig:density_evolution}. 
The upper panel of Figure \ref{fig:density_evolution} shows that the wide jets of the first pair compress some of the core material to form a dense shell, with which the narrow jets interact, as the later panels show. 
The narrow jet of the second pair propagating downwards is narrower than the one propagating upward. Each jet inflates an ear in the early stages (Figure \ref{fig:density_evolution_2}). The differences between the two ears are small. The jets then break out to form a nozzle, as shown in the lower panel of Figure \ref{fig:density_evolution_2} for the down jet. The dense volume around each nozzle and on the dense shell is a ring, a circum-jet ring (yellow-colored zones in the lower panel of Figure \ref{fig:density_evolution_2}). The jets form these rings by pushing the material of the dense shell to the sides. The shaping of these rings, which form morphologies similar to those observed in some CCSNRs, is the main result of this study.       

\begin{figure}
\centering
\includegraphics[trim=1.1cm 0cm 0.0cm 0.0cm ,clip, scale=0.28]{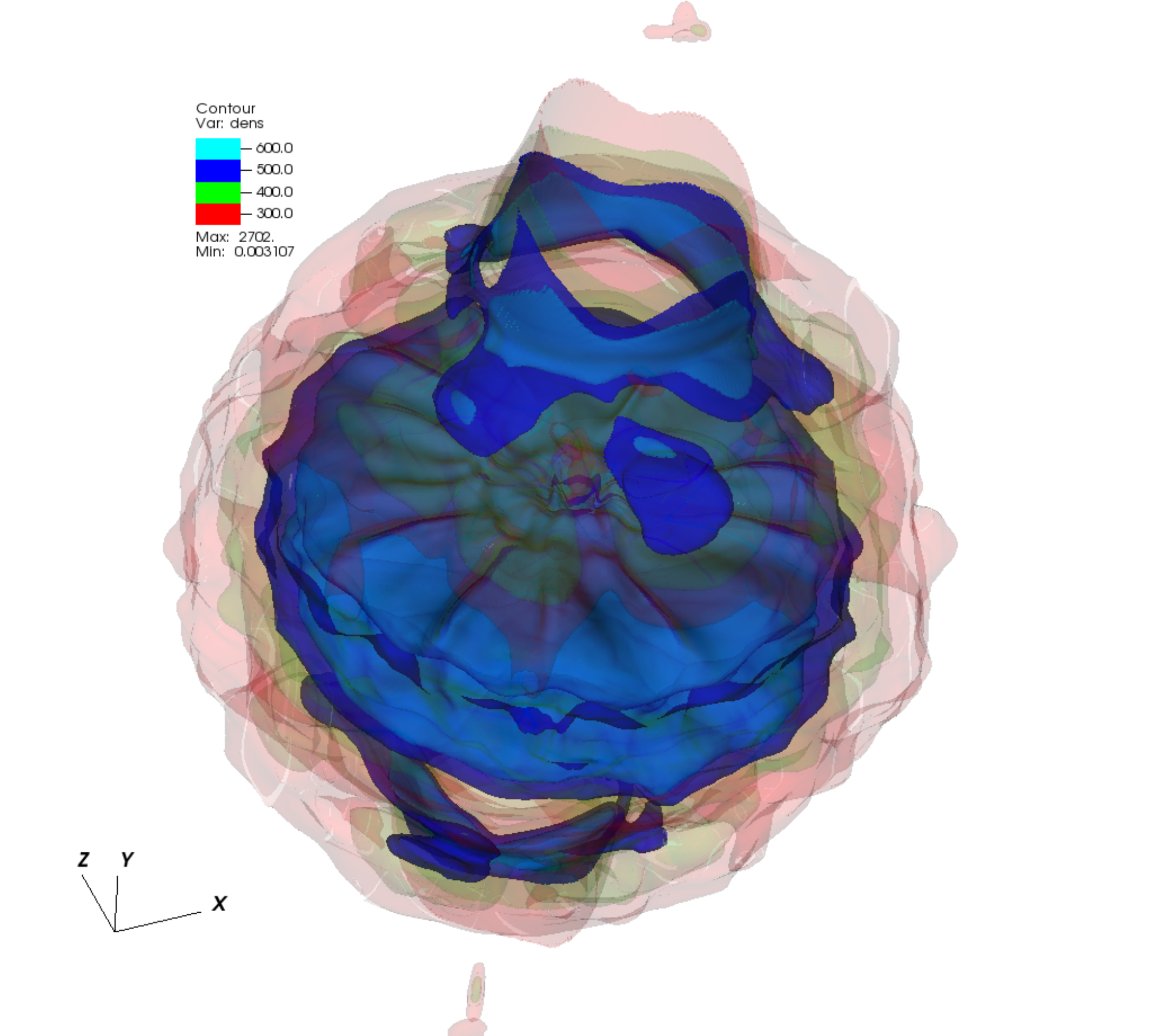} 
\caption{Three-dimensional visualization of the gas density for Run A at $t=6 \s$, composed of four equidensity surfaces as the color bar shows in units of $\g \cm^{-3}$. The viewing direction is at $56^\circ$ to the symmetry axis of the two narrow jets, which is the symmetry axis of the pair of wide jets. The figure emphasizes the formation of dense rings, which arise from the interaction between the jets and the core material. The up ring is the wavy, dark, and bright blue ring on top of the blue shell.   }
\label{fig:3D}
\end{figure}

In Figure \ref{fig:3D}, we present a 3D view of the density at $t=6 \s$. The observer here is at $56^\circ$ to the symmetry axes of the jets/rings. Four equidensity surfaces form this image, as indicated. The up ring is clearly visible as a dark and bright blue ring. Part of the down ring is also visible. The outer red dots are the material along the jet axis. Instabilities, such as the Rayleigh-Taylor instability, make the ring clumpy and distorted. The finite resolution of the numerical grid seeds the instabilities. Here, the ring appears as a wave with four peaks.  
%

\begin{figure}
\centering
\includegraphics[trim=0.0cm 0cm 0.0cm 0.0cm ,clip, scale=0.19]{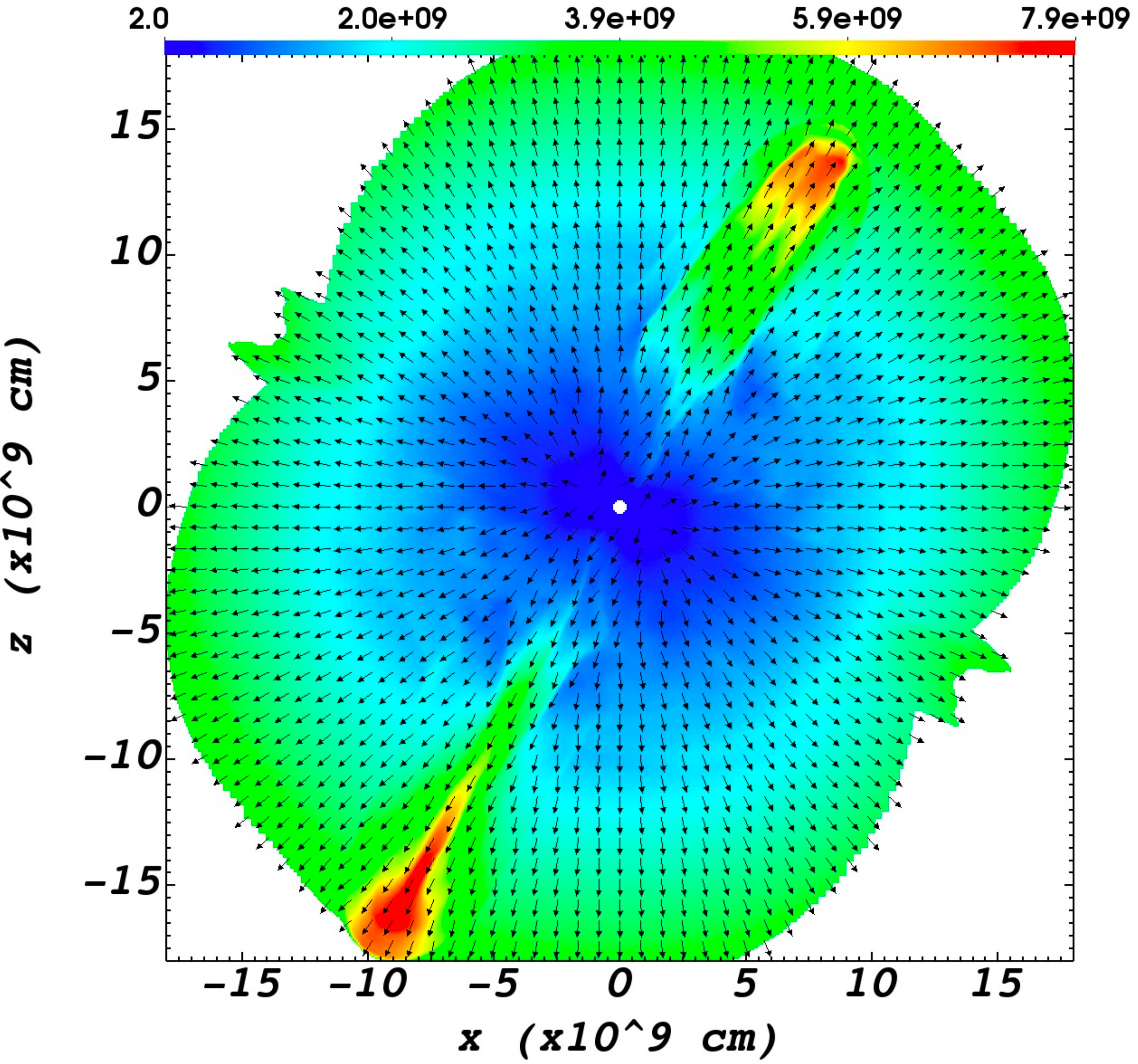} 
\caption{Velocity map in the meridional $xz$ plane for Run A at $t=6 \s$, with arrows indicating the direction and color the magnitude according to the color-bar in $\cm \s^{-1}$. Only regions with density of $\rho>5 \g \cm^{-3}$ are shown. The forward shock has exited the numerical grid at $t \simeq 3 \s$}
\label{fig:velocity_map}
\end{figure}

In Figure~\ref{fig:velocity_map}, we present the velocity in the meridional plane at $t=6 \s$, but remove the very low density gas, $\rho < 5 \g \cm^{-3}$, around the ejecta as these regions might suffer numerical artifacts; color depicts the velocity magnitude in $\cm \s^{-1}$ and arrows the direction.  Note that the forward shock has exited the numerical grid at $t\simeq 3 \s$.  The velocity map emphasizes the following. (1) The very fast velocities we obtain for the outer ejecta are because we simulate a very energetic explosion. (2) The very fast-moving (but not relativistic) fronts of the jets. (3) Although the first pair half-opening angle was $60^\circ$, it still accelerates material to high velocities even in the equatorial plane. 

\begin{figure}
\centering
\includegraphics[width=\linewidth, trim=4.5cm 0cm 4.5cm 0cm, clip]{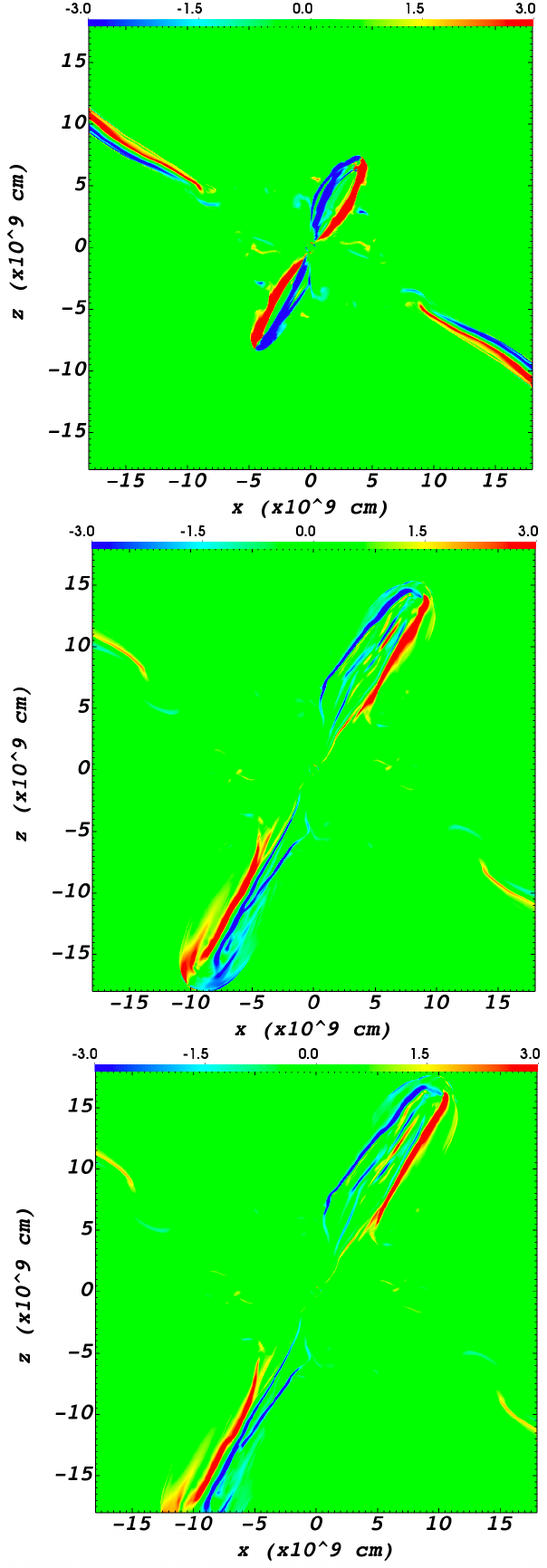}
\caption{ The vorticity component perpendicular to the plane, $(\nabla \times \vec{v})_y$, at three times, $t=4.4$, $6$, and $6.4 \s$ of Run A. The color bar runs from $-3 \s^{-1}$ (deep blue; counterclockwise) to $3 \s^{-1}$ (deep red; clockwise). }
\label{fig:curl_evolution}
\end{figure}

There is a strong shear as the jets expand and inflate the ears. The shear excites vorticity in the flow. We present the vorticity, $(\nabla \times \vec{v})_y$, in Figure~\ref{fig:curl_evolution}, in the $xz$ meridional plane and at three times and in units of $\s^{-1}$. The figure shows that the time scale of the vorticity is shorter than the flow time $(\nabla \times \vec{v})^{-1} < t$; in narrower regions, the vorticity reaches values of $(\nabla \times \vec{v})_y > 20 \s^{-1}$. This implies that mixing is very efficient within that region. However, Figure \ref{fig:curl_evolution} shows that the jets induce vortices only in a thin envelope around the jets; only there the jet's material and the core material mix by the jet-induced turbulence. The vorticity acting together with the shear can substantially amplify magnetic fields in these regions, in a process termed jet-driven dynamo (JEDD), and which was studied for jets in other environments, like active-galactic nucleus jets in clusters of galaxies (\citealt{Soker2017JEDD}; later, \cite{Tripathietal2026} adapted this process and developed a version of JEDD in a more rigorous manner).  Another effect of the velocity shear is the development of the Kelvin-Helmholtz instability. The thin, faint filaments inside the lobes in the two lower panels of Figure \ref{fig:density_evolution_2} are along the high vorticity regions. They are not straight; we attribute their wiggling to the Kelvin-Helmholtz instability and the vorticity.   

\begin{figure}
\centering
\includegraphics[width=\linewidth, trim=4.5cm 1.5cm 4.5cm 0cm, clip]{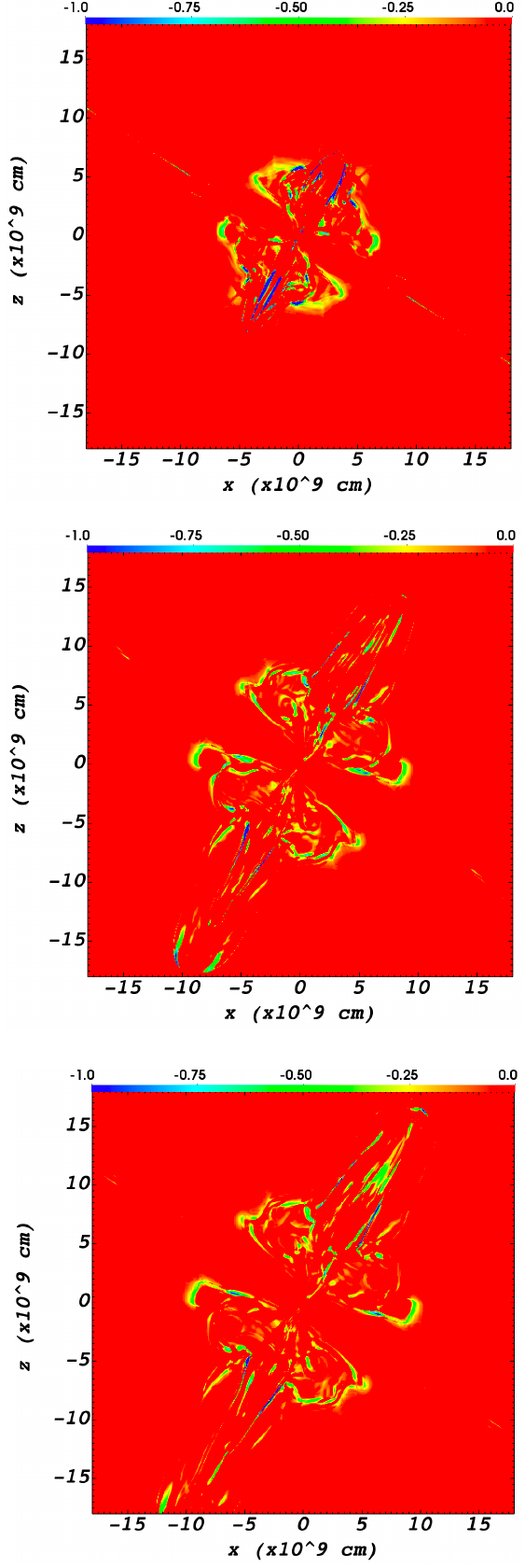}
\caption{The Rayleigh--Taylor instability diagnostic $f_{\rm RT}$ (equation \ref{eq:RTinstability}) for Run A at $t=4.4$, $6$, and $6.4 \s$. Negative values indicate RT-unstable regions, while red regions are stable.}
\label{fig:rt_evolution}
\end{figure}

The interaction of the two pairs of jets with the core material inflates high-temperature, low-density gas that accelerates the core material and the CSM. This interaction is prone to the Rayleigh-Taylor instability. To quantify the development of Rayleigh--Taylor (RT) instabilities, we use the same diagnostic quantity as in \cite{Braudoetal2025} 
\begin{equation}
f_{\rm RT} \equiv \frac{1}{\rho} \sqrt{\left| \vec{\nabla}P \cdot \vec{\nabla}\rho \right|} \,
\mathrm{sgn}(\vec{\nabla}P \cdot \vec{\nabla}\rho), 
\label{eq:RTinstability}
\end{equation}
where $P$ is the pressure and ${\rm sgn} (\vec{\nabla}P\cdot\vec{\nabla}\rho)$ is the sign of the scalar product of pressure and density gradients. 
If $f_{RT}$ is negative, the region is unstable with a typical growth rate of $-f_{RT}$, and a typical growth time of $-1/f_{RT}$.
Figure~\ref{fig:rt_evolution} presents the time evolution of $f_{\rm RT}$ at $t=4.4$, $6$, and $6.4 \s$.
The red region is stable. The green and blue regions have a growth time of $\simeq 1-2 \s$. This timescale is an order of magnitude shorter than the flow time, indicating that these instabilities have time to grow. 
The instabilities enhance mixing between the jets' material (both pairs of jets) and the core material, but again only in a thin region around the jets and the ears or bubbles they inflate. 

\subsection{Emission integral maps}
\label{seubsec:EI}

We aim to reveal pairs of rings and compare them with observations. The emission integral (EI), as given in equation (\ref{eq:EI}), is the numerical quantity closest to observations we can calculate from our simulations to draw relative intensity maps; we take $Y_s$ as the line-of-sight coordinate.  
 The emission integral, which is the integral of the emission measure ($\propto \rho^2$), gives the approximate synthesis observed map in the X-ray. It might be only qualitatively correct for synchrotron radio emission, because the emission depends sensitively on the number density of accelerated electrons and the magnitude and orientation of the magnetic fields. The acceleration of electrons depends on the shock. In this study, we focus on the formation of rings, which we present in the 2D (Figures \ref{fig:density_evolution} and \ref{fig:density_evolution_2}) and 3D (Figure \ref{fig:3D}) maps. For these, we do not claim a perfect fit between our simulations and observations, nor in X-ray, nor in other wavelengths. We only emphasize the formation of rings. 

As we explain in Section \ref{sec:Numerics}, we do not accurately simulate the inner, dense equatorial gas because we do not launch later jets (which we leave for a future study). Therefore, to allow showing the outer regions we are focusing on, we removed the dense gas from the emission integral maps. The inner dense region will affect viewing angles close to the polar axis of the jets; therefore, we do not present images at small inclination angles.

\begin{figure}
\centering
\includegraphics[width=\linewidth, trim=0.cm 0.0cm 0.cm 0cm, clip]{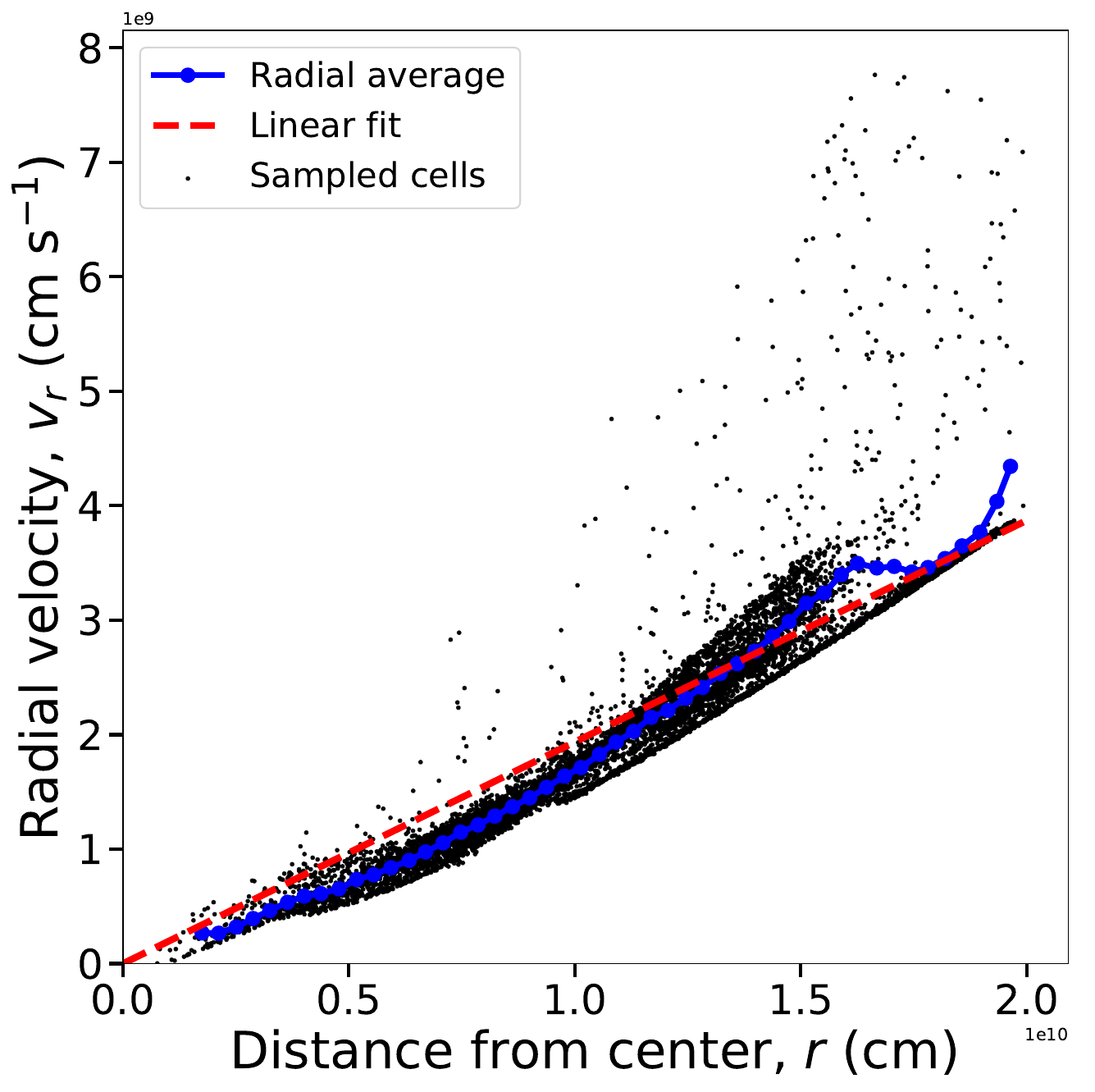}
\caption{  The velocity versus the distance at $t=6 \s$ for a random set of 10,000 grid cells having a density of $\rho > 5 \g \cm^{-3}$. The blue line is the average velocity of the points, and the dashed-red line is a linear fit through the origin. This graph shows that the ejecta is close to homologous expansion, but not yet perfectly so.  
}
\label{fig:HomologousExpansion}
\end{figure}

At $t\gtrsim 6 \s$, the kinetic energy of the material we show in the figures to follow is about 8 times the thermal energy. This implies that the expansion is nearly homologous, and the outer ejecta we study will maintain their structure until they collide with a massive CSM or ISM.  In Figure \ref{fig:HomologousExpansion} we present the velocity versus distance at $ t =6 \s$ for $10^4$ points chosen from many locations in the ejecta. We also show the average velocity of the points, and a linear fit through the center. Although not perfect, the outflow is nearly homogeneous. 

\begin{figure}
\centering
\includegraphics[width=\linewidth, trim=0.cm 3.5cm 0.cm 0cm, clip]{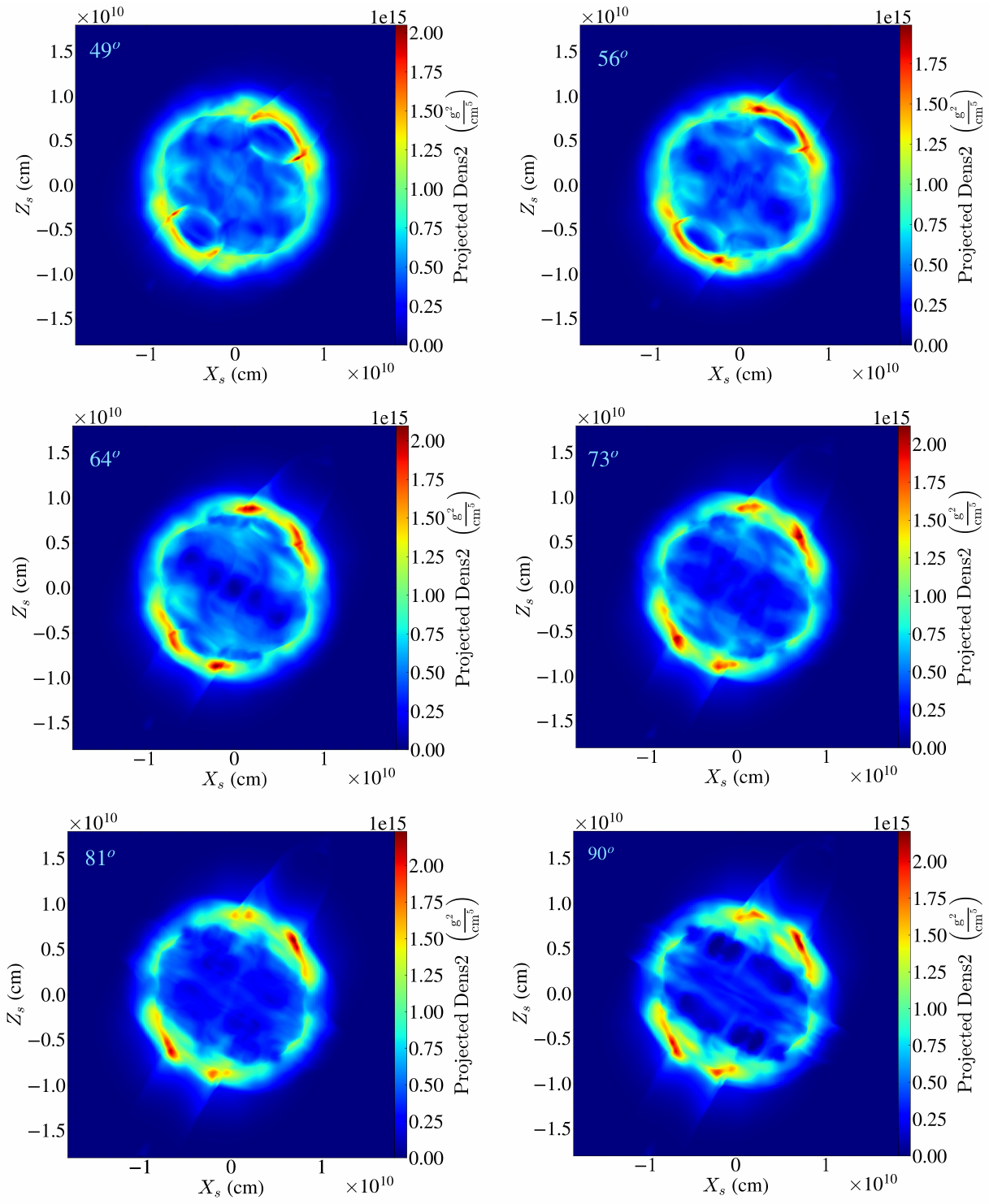}
\caption{Emission integral (equation \ref{eq:EI}) maps for Run A at $t=6.4 \s$ for six different inclination angles as indicated (the angle of the narrow-jet axis to the line of sight). The rings are on the upper right and lower left of the dense shell in each panel. They are easy to notice in the upper panels as a bright ring around a very faint (deep blue) zone.  }
\label{fig:em_angles_A}
\end{figure}

Figure \ref{fig:em_angles_A} shows the emission integral maps of Run A at $t=6.4 \s$ for six viewing angles as indicated, where the angle is from the symmetry axis of the narrow jets to the line of sight. We recall that the down jet is narrower than the up jet, so the two polar directions should not be identical. 

From Figure \ref{fig:em_angles_A}, we learn the following: 
(1) At large inclinations $60^\circ \lesssim i \le 90^\circ$, each ring appears as two bright, thin, and short bright zones along the dense and bright shell, pointing outward toward the symmetry axis. There are some fainter filaments between the two bright zones, which we elaborate on in Section \ref{sec:G46}. (2) The rings start to appear at an inclination of $\simeq 60^\circ$ and below. We further elaborate on the rings in Section \ref{sec:Rings}. (3) Although Run A was set to have axisymmetry around the common axis of the two pairs of jets, there are small differences on the two sides of the symmetry axis of each ring. This is due to the numerical grid's limited resolution. The Cartesian grid introduces perturbations that are not axisymmetric because the angle of the jet axis to the $z$-axis is $30^\circ$ and not $45^\circ$. These differences do not change our conclusions, but should be kept in mind (we have no computer resources to increase the resolution in the present setting).  (4) Although hard to notice, there is an ear protruding from each of the two rings. In observations, in many cases these will be below the detection limit, as the jets have already broken to large distances. 

\begin{figure}
\centering
\includegraphics[width=\linewidth, trim=0cm 3.5cm 0cm 0cm, clip]{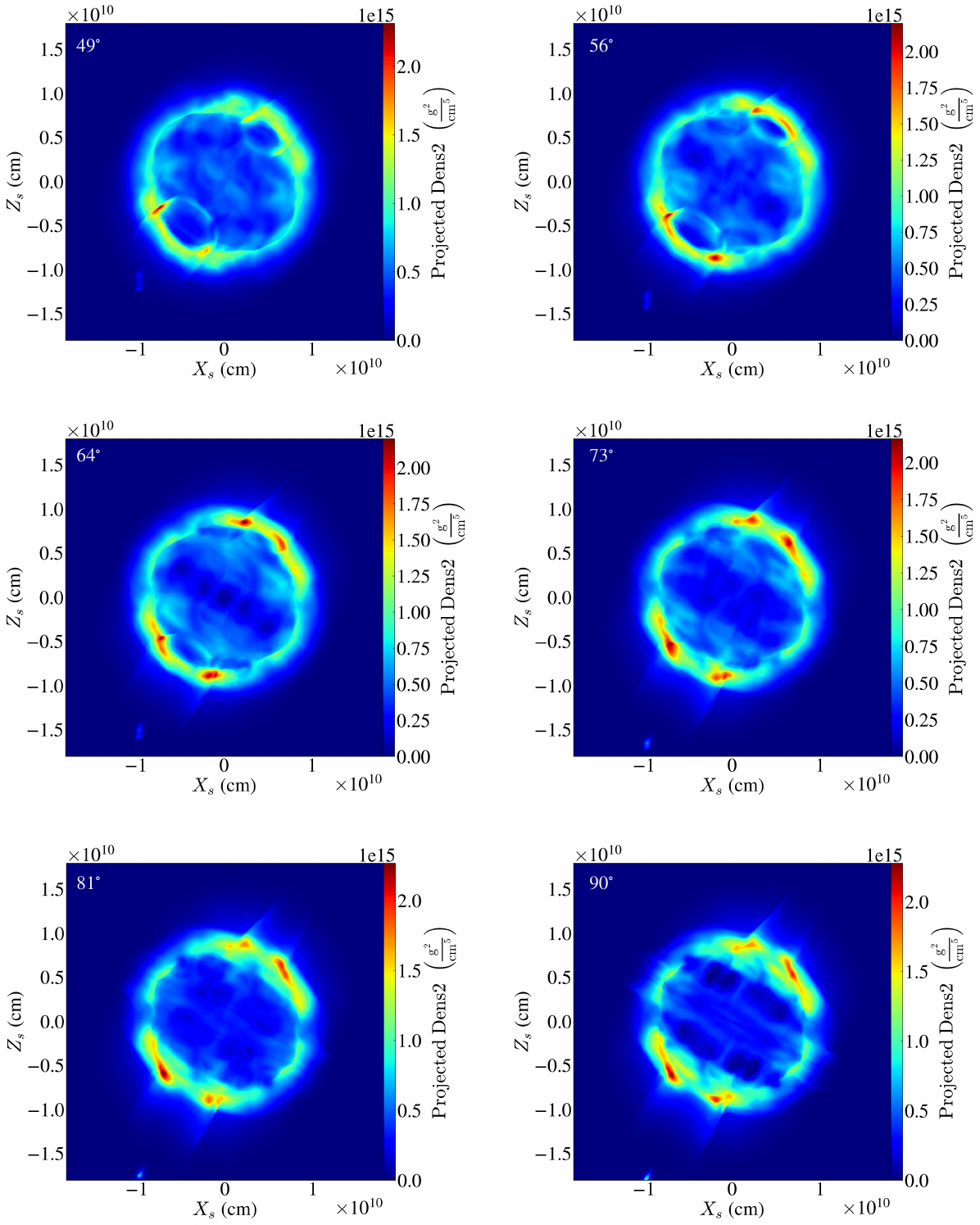}
\caption{Similar to Figure \ref{fig:em_angles_A} and for the same viewing angles, but for Run A2, where the up jet is third the power and the down jet is three times the power of the respective jets in Run A of Figure \ref{fig:em_angles_A}. Note that the color bar has a different scaling. 
}
\label{fig:em_run_A2}
\end{figure}

\begin{figure}
\centering
\includegraphics[width=\linewidth, trim=0cm 3.cm 0cm 0cm, clip]{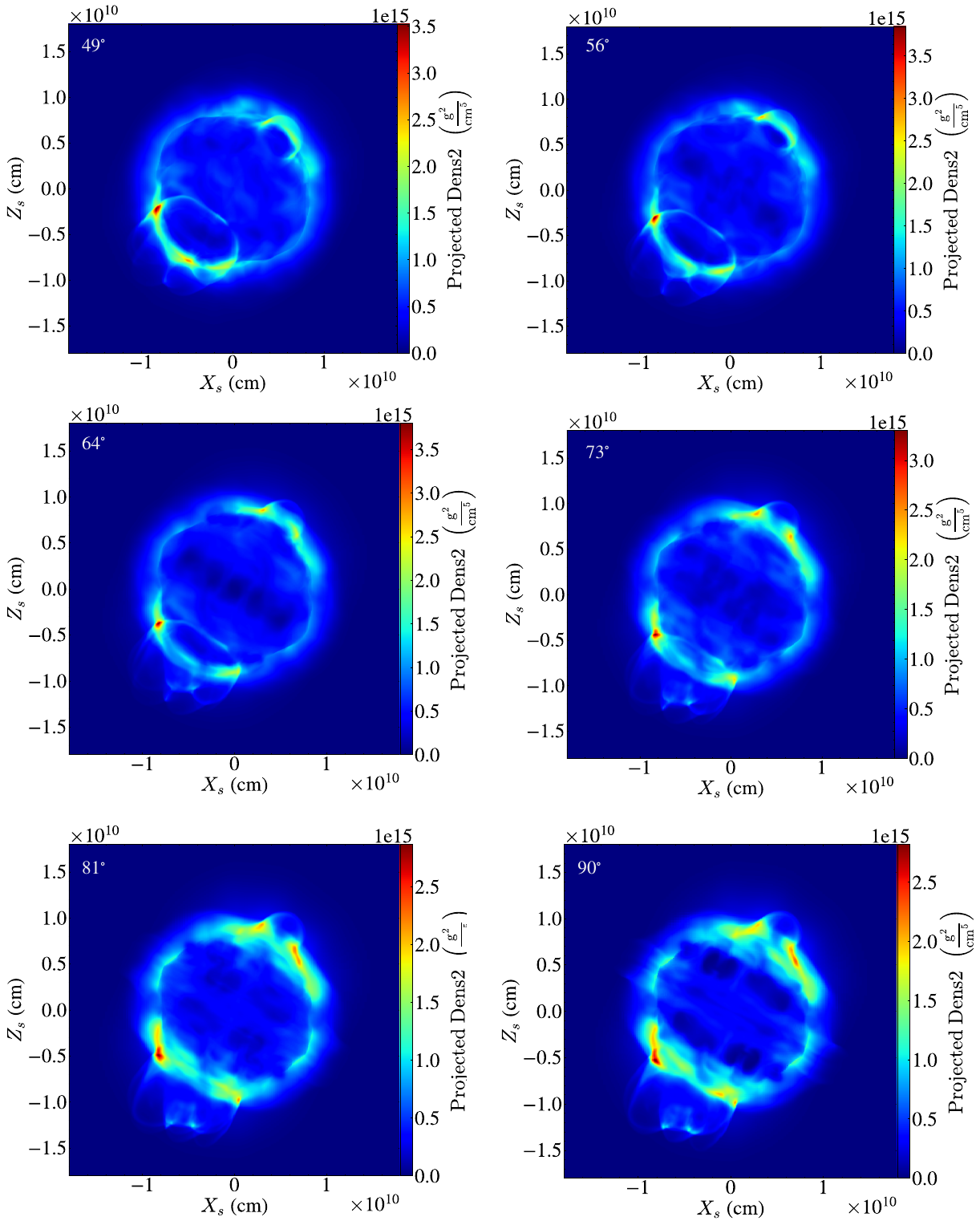}
\caption{ Similar to Figure \ref{fig:em_run_A2} but for Run A3, where the half opening angle of the two jets is $20^\circ$, instead of $10^\circ$ (up) and $5^\circ$ (down). Note that the color bars have different scalings. 
}
\label{fig:em_run_A3}
\end{figure}

We examine the sensitivity of the morphologies to some of the narrow jets' parameters. In Figure \ref{fig:em_run_A2} we present Run A2, where the up jet has a third of the energy (third of the mass) of the jets in Run A, and the down jet has three times the energy. 
We clearly see that the up ring is much smaller than the down ring. As the power of the jet increases, at least for the same speed, the size of the ring increases. The morphology changes little, even for an order-of-magnitude change in jet energy.  We already saw that the two jets of Run A, which differ by their half-opening angle, shape very similar rings. We further explore the influence of the half-opening angle in Run A3, which has the same jets' energies as in Run A2, but both jets have a half-opening angle of $20^\circ$. We present the emission integral of Run A3 in Figure \ref{fig:em_run_A3}. The wider jets have lower ram pressures and take longer to break out. Therefore, we leave the full exploration of these cases for future study. Here, we note that qualitatively, the rings of Run A3 and their morphologies are similar to Run A2. There are differences, though. The less energetic jet (up jet) inflates a small ear, and the ring is less dense and less pronounced than the up ring in Run A2. The up ring in Run A3 is of a similar size to that in Run A2. The ear that the down jet inflates has a complicated structure, which we do not explore here. The down ring of Run A3 is noticeably larger than the down ring of Run A2. 

\begin{figure}
\centering
\includegraphics[width=\linewidth, trim= 0cm 3.cm 0cm 0cm, clip]{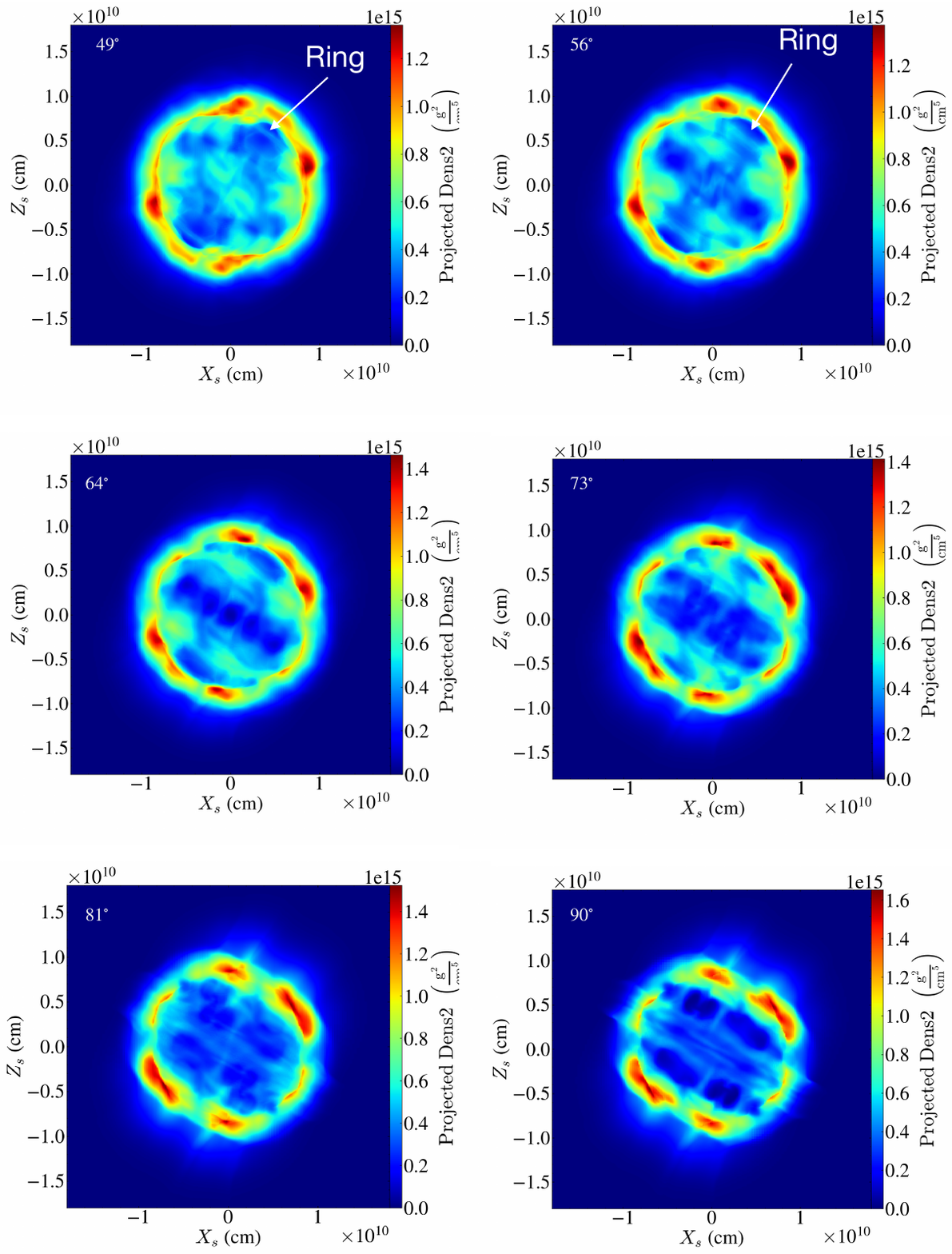}
\caption{Emission measure maps for Run AEi at $t=6.4 \s$, which differs from Run A (Figure \ref{fig:em_angles_A}) in the start of the second pair of jets being $t_2=1 \s$ instead of $t_2=3 \s$. 
}
\label{fig:em_run_AEi}
\end{figure}

We simulated two cases with an earlier second pair of jets: Run AEi, where we start launching the second pair at $t_2=1 \s$, and Run AEii, where $t_2=2 \s$; in both cases, the launching lasts for one second. We present the emission integral of Run AEi in Figures \ref{fig:em_run_AEi}.  The narrow jets do form two polar circum-jet rings, but their circumferences are fainter. We point at the up ring in the two upper panels. The rings of Run AEii (that we do not present here) have an appearance between Run A and Run AEi. We conclude (also based on trial simulations we do not mention here) that jets form more prominent circum-jet rings when the dense shell they interact with is more developed, namely denser and thinner. This might suggest that in CCSNRs with large and clear rings, there was at least one pair of jets at a late time during the explosion, $t_2 \gtrsim 2 \s$.  

\begin{figure}
\centering
\includegraphics[width=\linewidth, trim= 0cm 3.cm 0cm 0cm, clip]{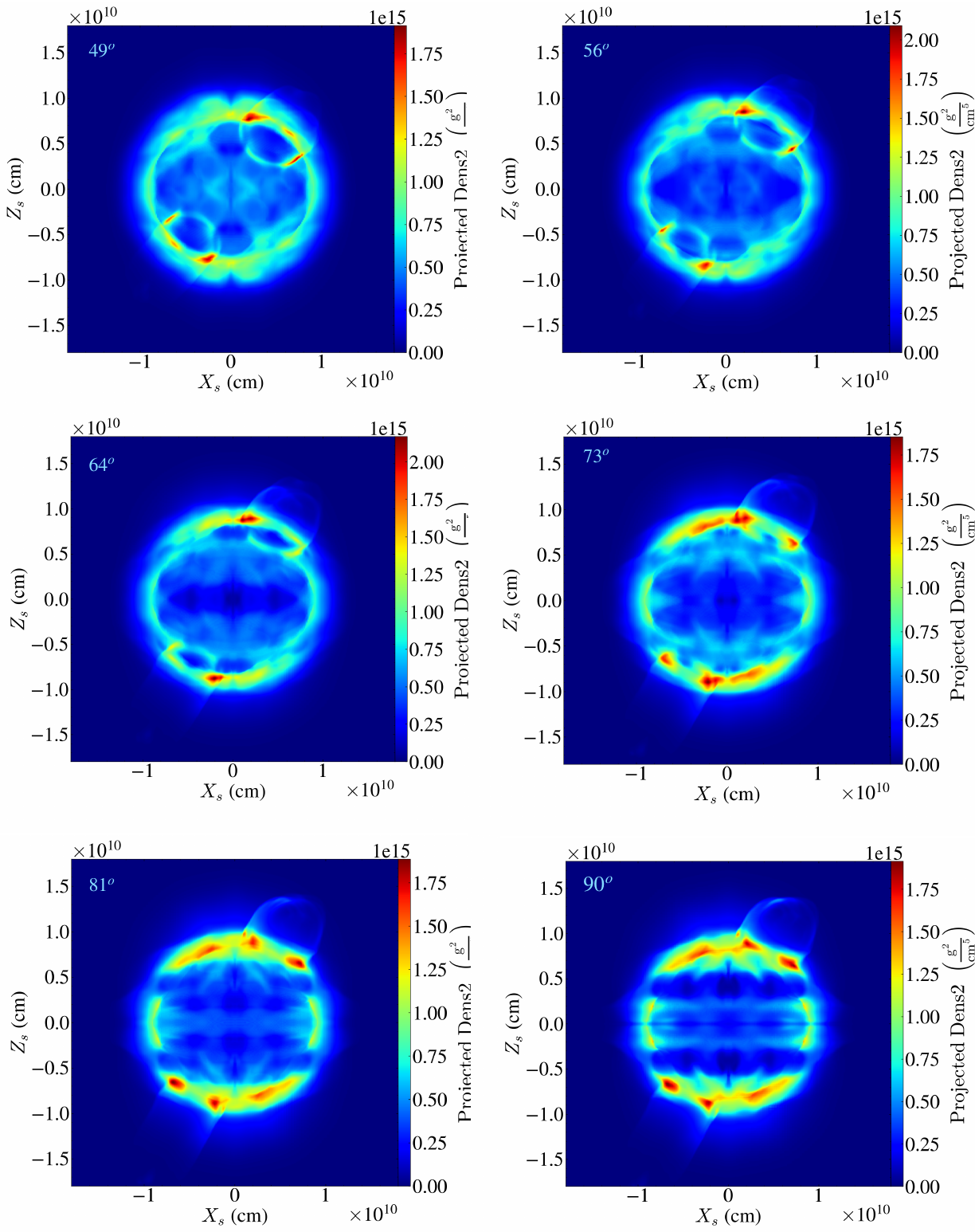}
\caption{Emission measure maps for Run M, where the axis of the first pair is along the $z$ axis, which is $30^\circ$ from the second pair of jets, at $t=6.4~\mathrm{s}$.
}
\label{fig:em_run_M}
\end{figure}

In Figure \ref{fig:em_run_M}, we present the results for Run M, where the axis of the first pair is along the $z$-axis, implying that the axis of the second pair, the narrow jets, is tilted by $30^\circ$ to that of the wide jets. The main difference between simulations, Run A and Run M, is that in Run M, the two sides of each ring are much more asymmetrical. This is expected because of the misalignment. This outcome is important in comparing our results to observations, as we discuss in Section \ref{sec:G46}.   

 We note that the tails of the narrow jets interact with the dense shell near the stellar surface. This raises the question of whether, in the case of a hydrogen-rich star, i.e., a red-supergiant star, the rings will lose their identity as they expand through the large hydrogen envelope. \cite{Braudoetal2026} use the same stellar model, but launch the jets much deeper in the core and within a much shorter time. Their jets are choked deep in the core. Still, the structures that the jets form deep in the core maintain their identity long after breakout. The reason is the steep density gradient. The jets form structures by compressing inner core material to higher densities. The very high-density structures interact with the much lower stellar material further out. This interaction changes the structure somewhat, but does not destroy it. Therefore, we expect that the dense rings we obtain here will survive the interaction with a low-density hydrogen-rich stellar envelope in cases where the shaping jets are energetic, and the hydrogen-rich envelope is not very massive. The core material has a higher atomic weight than the envelope, i.e., it can be S-, Si-, and O-rich. This ensures it maintains a higher density. Although our current computer resources do not allow us to simulate a larger grid or the expansion of the system in a large hydrogen-rich envelope, future simulations must study this process.

Overall, these comparisons demonstrate that the formation of circum-jet rings is a  possible outcome across a large volume of the parameter space. For the set of simplifications we used, using a WR-type stellar model without an extended hydrogen-rich envelope, not including a dense CSM, and setting an explosion with a very energetic pair of wide jets before launching the narrow jets, the formation of pairs of rings does not require fine-tuning. Future studies should relax our assumptions. 

\section{The CCSNR G46.8-0.3}
\label{sec:G46}

\begin{figure}[]
	\begin{center}
\includegraphics[trim=0.0cm 3.5cm 0.0cm 0.0cm ,clip, scale=0.7]{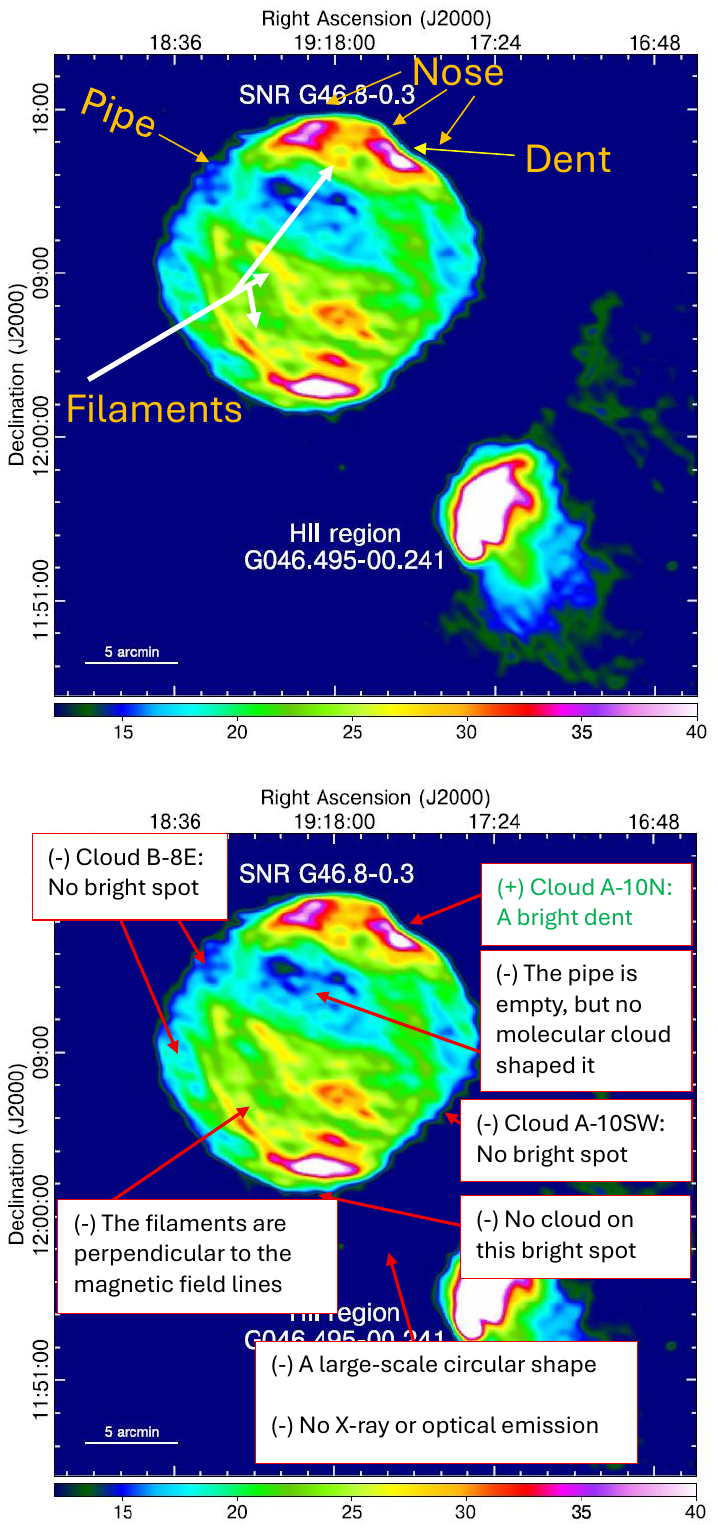} 
\caption{ A radio image of SNR G46.8–0.3 at 1.4 GHz adapted from \cite{Supanetal2022}. The color scale is linear in mJy/beam. The white bar marks 5 arcmin.
Upper panel: We mark four morphological features. The pipe: The empty (faint) zone stretching from the northeast to the west. An HI emission inside it \citep{Supanetal2022} shows that it is transparent, hence of low density. The nose: The entire structure north of the pipe, with two bright spots. The dent, on the outer boundary, coincides with a bright zone. The filaments: The bright stripes, shaped as arcs, to the south of the pipe.   
The HII region to the southwest of the CCSNR is unrelated to this CCSNR. 
Lower panel: We list one observational feature compatible with molecular-cloud shaping (marked with a plus sign) and several that are not (marked with a minus sign). }
\label{Fig:SNRG46FigureRadio}
\end{center}
\end{figure}

Several papers present radio images of SNR G46.8–0.3 (e.g., \citealt{Dubneretal1996, RanasingheLeahy2018, Shanahanetal2023}). We adapt an image from \cite{Supanetal2022}, which we present in Figure \ref{Fig:SNRG46FigureRadio}. 
\cite{Supanetal2022} study SNR G46.8–0.3 and its interaction with the ambient gas. They estimate the age of this CCSNR to be $\approx 10^4 \yr$. They conclude that SNR G46.8–0.3 does not interact with the neutral HI region but does interact with the molecular clouds. We argue that although SNR G46.8–0.3 interacts with the molecular clouds, there are morphological features that this interaction does not explain. 

The reasons that the interaction with the molecular clouds is not sufficiently strong to account for all morphological features are as follows, and as we list in the lower panel of Figure \ref{Fig:SNRG46FigureRadio}. 
There is, however, one morphological feature that is compatible with a strong interaction. The cloud that \cite{Supanetal2022} term A-10N in their figure 7 is outside the dent and the bright spot on the northwest; we mark this with `(+)' in the lower panel of Figure \ref{Fig:SNRG46FigureRadio} to indicate that it is compatible with shaping by the molecular cloud.  
 On the other hand, some morphological features do not support shaping by the molecular clouds. 
(1) There are no detections of X-ray and optical emissions from SNR G46.8–0.3. This might indicate that there are no strong shocks that ionize the molecular clouds. 
(2) This CCSNR has a large-scale circular structure, as \cite{Supanetal2022}  noted. It was not distorted much by the molecular clouds that have a highly non-spherical distribution around SNR G46.8–0.3  (figure 7 in \citealt{Supanetal2022}). 
(3) The cloud that \cite{Supanetal2022} mark A-10SW in their figure 7, is on the southwest edge of the CCSNR, but there is no strong radio emission there. Namely, there is no indication of a strong interaction there. 
(4)  The B-8E cloud does not form a bright SNR region in its interaction location on the northeast. 
(5) No molecular cloud on the edge of the south bright region. 
(6) There is no molecular cloud that seems to have shaped the empty zone we term `pipe.' The presence of HI emission within the pipe \citep{Supanetal2022} suggests that it lies behind the CCSNR and that the pipe is a low-density region.  
(7) The arc-shaped filaments stretched perpendicular to the magnetic field lines, not along them. \cite{Velazquezetal2025} and \cite{VelazquezReynoso2025}, among others, show that the interaction of SNRs with a magnetized ISM forms dense zones along the magnetic field lines. The filaments in the southern half of SNR G46.8–0.3 extend in the general direction from northeast to southwest. However, the magnetic field lines' general direction in SNR G46.8–0.3 is from northwest to southeast \citep{Sunetal2011, Shanahanetal2022}. 

\begin{figure}[]
	\begin{center}
\includegraphics[trim=0.0cm 8.5cm 0.0cm 0.0cm ,clip, scale=0.7]{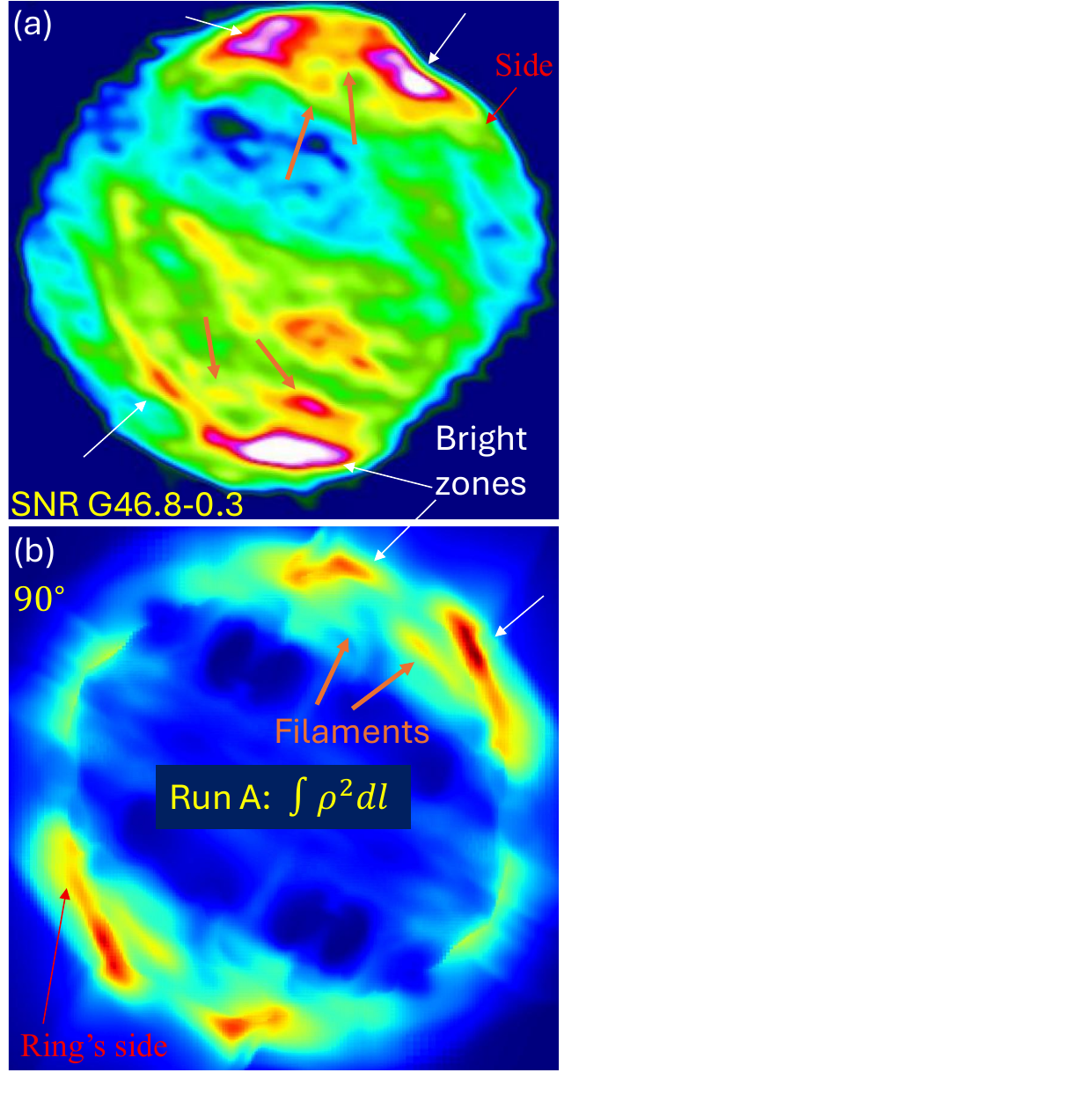} 
\caption{ Comparing the outer bright regions in the radio image of SNR G46.8–0.3 at 1.4 GHz \citep{Supanetal2022} with the emission integral at $90^\circ$ of our Run A of a pair of narrow jets that interacts with a pair of wider jets. We point at the bright zones in the north and south (thin white arrows), the filaments between the bright zones on each side (thick orange arrows), and the ring's side (red arrow). The color coding of the two panels is not identical, though similar. In both the observed image and the simulation, the peak emission of the bright zones is $\simeq 1.5$ times that of the filaments.  
}
\label{Fig:SNRG46FigureCompare}
\end{center}
\end{figure}

We accept that SNR G46.8–0.3 interacts with the molecular clouds, but it seems that the CCSNR shapes the clouds more than the clouds shape the CCSNR. The interaction is responsible for the radio emission. 
We raise the possibility that the main structure of the `nose' is due to jets, or more precisely, the interaction of a narrow jet with a wider one. Counter-jets are responsible for the bright regions in the south. 
In Figure \ref{Fig:SNRG46FigureCompare} we present the same radio image of SNR G46.8–0.3 as in Figure \ref{Fig:SNRG46FigureRadio} at 1.4 GHz \citep{Supanetal2022}, and an emission integral figure from our Run A at a projected angle of $90^\circ$ (as in Figure \ref{fig:em_angles_A}). We note that our simulation of a narrow jet interacting with a wider jet can reproduce the general structure of the `nose': (1) Two bright zones on the two sides of the symmetry axis of the jets (pointed at by thin white arrows). Each pair of bright zones is the projected dense zones of a circum-jet ring. We also point with a red arrow on the side of the rings, an extension from the ring. (2) Fainter filaments in the region between the bright zones (pointed at by thick orange arrows).  We suggest that jet-jet interaction formed each of the two opposite pairs of bright zones of SNR G46.8–0.3. The greater number of filaments in the south might hint at more interacting jets.  
 
We note that, although we construct cylindrically symmetric jets in our simulation Run A, the two bright zones of the same ring are not equal. This is because of the finite numerical resolution of the Cartesian grid. Since we inject the jets not along any grid axes, nor at $45^\circ$, the numerical grid introduces unequal properties around the jet axis. Although here the effect is numerical in Run A, in reality, we expect instabilities, precession, jittering, magnetic fields, and above all misalignment to introduce such effects. In Run M, where the second pair of jets is misaligned with the first, the ring side is much larger than in Run A (Figure \ref{fig:em_run_M}). Therefore, unequal sides in the projection of the SNR on the plane of the sky are expected. 

 We end with listing some limitations of our simulations in comparison with SNR G46.8–0.3. 
(1) Our simulations are for a striped-envelope progenitor. We do not include the propagation of the structure through the helium- and hydrogen-rich envelope layers of a type II CCSN progenitor. (2) We did not include interaction with a CSM and the ISM. This SNR interacted with some ISM as we mentioned above. This interaction adds mass to the expanding ejecta and excites shocks within it, processes that alter its structure. (3) Our synthesis maps are emission integral maps that mimic thermal X-ray emission, but not radio emission. In this old CCSNR, the reverse shock has passed through the core, and the ejecta have cooled; we expect no strong X-ray emission.  The radio maps of SNR G46.8–0.3 show synchrotron emission from accelerated electrons and magnetic fields; the radio emission depends on the magnitude and direction of the magnetic fields. We do not include these in our analysis. For these factors, the comparison of our simulations with the observations of SNR G46.8–0.3 is qualitative, and not quantitative. Our simulations should motivate more accurate simulations that include these processes.

The main conclusion of this section is that jets shaped the filaments and bright zones in the north and south of SNR G46.8–0.3 (in the north, they form the `nose'). Interaction with the molecular clouds cannot explain these morphological features (lower panel of Figure \ref{Fig:SNRG46FigureRadio}), while the simulations of interacting jets can reproduce such morphological features (Figure \ref{Fig:SNRG46FigureCompare}). 

\section{Circum-jet rings}
\label{sec:Rings}

We obtain circum-jet rings, one ring around each narrow jet (Figures \ref{fig:em_angles_A} - \ref{fig:em_run_M}). We obtain such rings around CCSN jets in our previous study \citep{SokerAkashi2025} with different jet settings. It seems that many jet settings (but not all) can lead to the formation of circum-jet rings, though with different properties and appearances. This is in line with the non-negligible number of CCSNRs that exhibit morphological features of circum-jet rings and with their diverse morphologies. Presently, there are about 6 or 7 CCSNRs with claimed circum-jet rings out of about 20 CCSNRs with high-quality images and structures that were not destroyed by the ambient gas, and for which shaping by jets was identified.   
$\bullet$\cite{ShishkinKayeSoker2024} identified a partial ring (which they termed a cavity) in an image of the Cygnus Loop by \cite{Raymondetal2023}. 
$\bullet$ \cite{Soker2025SNRJ0450} identified a pair of opposite rings around a symmetry axis in an image of SNR J0450.4-7050 from  \cite{Smeatonetal2025SerAJ}. 
$\bullet$ \cite{SokerShishkinW49B} identified circum-jet rings in an HST image of SNR 0540-69.3 from \citet{Morseetal2006}.  A pair of opposite jets is seen in the 3D reconstruction by \cite{Larssonetal2021} of SNR 0540-69.3, and a pair of bubbles along the same axis in the 3D reconstruction by  \cite{Larssonetal2026}. 
$\bullet$ \cite{SokerShishkinW49B} identified rings in SNR W49B in images from \citet{Lopezetal2013a} (although it is not clear yet whether W49B is an SNR from a CCSN).
$\bullet$ A pair of opposite rings in SNR Circinus X-1 was robustly identified by \cite{Gasealahweetal2025}, who attributed the rings to post-explosion jets; in \cite{SokerAkashi2025}, we argued instead that the jets that shaped the rings participated in the explosion process in the framework of the JJEM.   
$\bullet$ \cite{SokerAkashi2025} identified a possible ring in SNR candidate G107.7-5.1 in an image by \cite{Fesenetal2024}. 
$\bullet$ \cite{Soker2025G11} identified pairs of opposite rings in the outer boundary of SNR G11.2-0.3 in an image by \cite{Robertsetal2003}. 

\begin{figure}[]
	\begin{center}
\includegraphics[trim=2.0cm 2.5cm 0.0cm 0.0cm ,clip, scale=0.64]{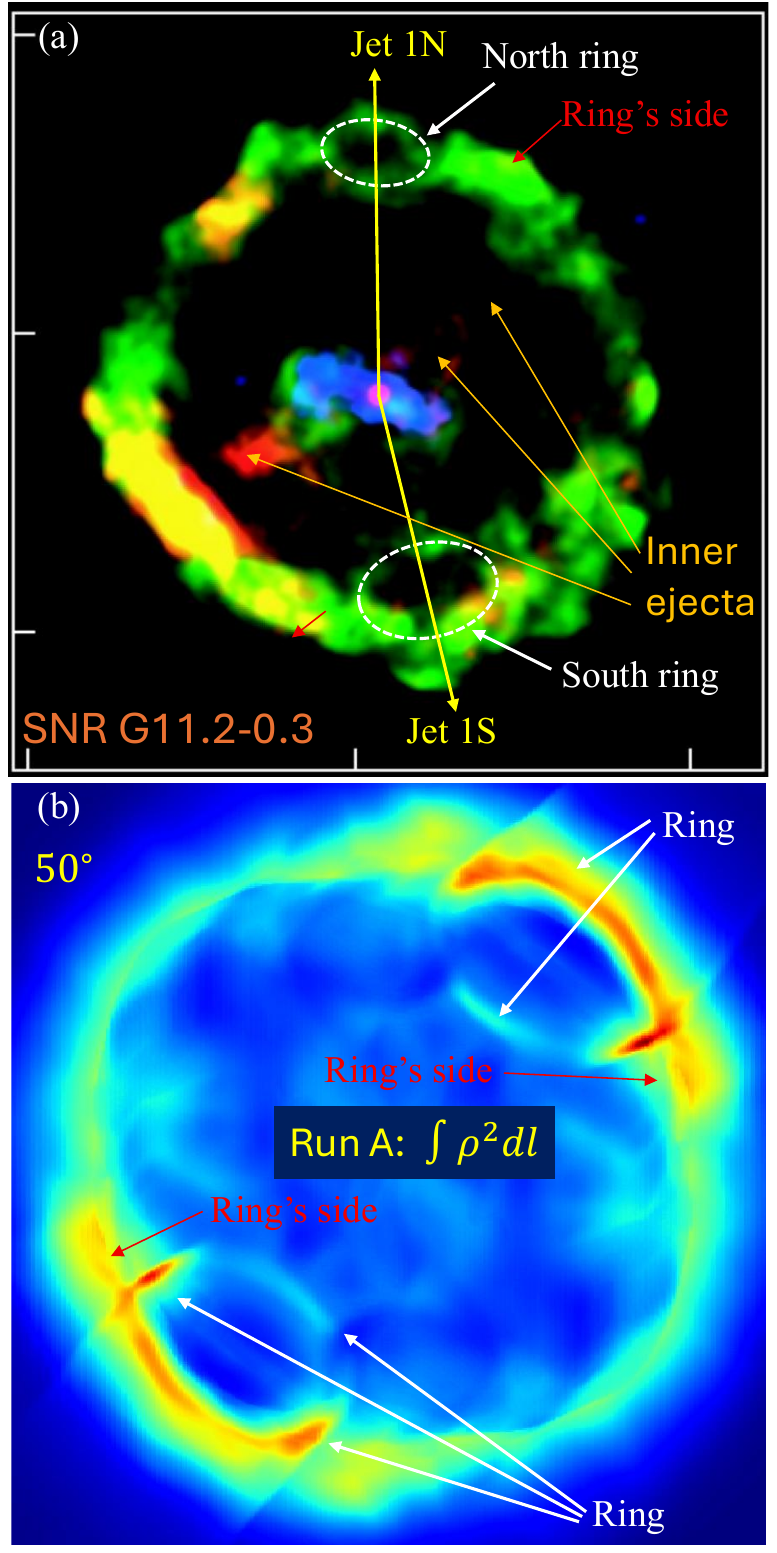} 
\caption{ (a) A figure adapted from \cite{Soker2025G11} of an image from  \cite{Robertsetal2003} that compares X-ray pulsar wind nebula emission and radio emission of SNR G11.2-0.3. Red: $0.6-1.65 \keV$ X-ray. Green: 3.5 cm radio. Blue: $4-9 \keV$ X-ray; all are at $5^{\prime \prime}$ resolution.
\cite{Soker2025G11} mark the identification of two rings and the directions of the two jets that shaped them. \cite{Soker2025G11} estimated inclination of the jets' axis to the line of sight to be $\simeq 50^\circ$. This gives the three-dimensional angle between the two jets of Pair 1 as $170 ^\circ$. 
Right ascension (J2000) ticks are 18:11:40, 18:11:30 and 18:11:20, and declination (J2000) ticks are $-19:27:00$, $-19:25:00$, and $-19:23:00$. We added the marks of the ring's side. 
(b) The emission integral of Run A at an inclination angle of $50^\circ$. 
 The emission integral better mimics X-ray than radio emission. Nonetheless, we note that the rings also appear in X-ray emission, although less prominently \citep{Soker2025G11}. The pulsar wind nebula is bound to an inner region elongated almost perpendicular to the rings, hence not likely to power the rings' emission.   
}
\label{Fig:SNRG46FigureG11Compare}
\end{center}
\end{figure}

From these seven SNRs, the morphologically closest to our simulations is SNR G11.2-0.3. In panel (a) of Figure \ref{Fig:SNRG46FigureG11Compare}, we present an image of this CCSNR, adapted from \cite{Soker2025G11}, who identified the rings in an image from \cite{Robertsetal2003}. \cite{Soker2025G11} identifies three symmetry axes, attributed to three pairs of energetic jets that operated during the explosion process in the framework of the JJEM. The rings are not exactly $180^\circ$ opposite, as observed in many other pairs of structural features in CCSNRs. Several processes might cause this bending, including instabilities that deflect the jets, unequal jets at launch (e.g., \citealt{Soker2024CounterJet}), and the core material deflecting the jets somewhat (e.g., \citealt{Gottliebetal2022}).    \cite{Soker2025G11} estimated the inclination angle of Pair 1 (jets' axis to line of sight) to be $\simeq 50^\circ$, and so in panel (b) we present the emission integral at $50^\circ$ from our simulation Run A. The simulations' rings of Run A are larger than those of SNR G11.2-0.3. However, lower energies lead to smaller rings; the low-energy (up) jet in Run A3 (Figure \ref{fig:em_run_A3}) shapes a ring that is $\simeq 83 \%$ the size of the rings in Run A that we present in Figure  \ref{Fig:SNRG46FigureG11Compare}. In any case, some morphological features suggest that the rings are indeed shaped by jets. We note these similarities between the image of SNR G11.2-0.3 and our simulation: (1) The rings are on the outer shell. (2) The outer part of the projected ring is along the shell, while the inner half protrudes inward. (3) In the southern ring of SNR G11.2-0.3, the outer half of the ring is much brighter than the inner half, as in our simulations. (4) There is an extended bright region to the side of the rings, which we term the ring's side. (5) There is material inside the outer shell of SNR G11.2-0.3, the inner ejecta, that was shaped by different pairs of jets. As mentioned, due to numerical limitations, we did not consider shaping the slower (inner) ejecta, but it does exist. \cite{Soker2025G11} suggested that the pair of jets that shaped the ring was the last one to be launched. It is possible that in the case of SNR G11.2-0.3, there were more pairs of jets between the first and last. The narrow jets we simulate here might have been the last.  

 We recall that in performing the synthesis observed map (the emission integral) that we present in panel (b) of figure \ref{Fig:SNRG46FigureG11Compare}, we removed the inner dense material. The reason is that in the present study, we are exploring the structure of the outer rings. In a follow-up paper, we \citep{AkashiSoker2026} launched another pair of jets, between the first and the third, which clear a large fraction of the central gas, leaving a bar. In that paper, we were aiming to reproduce the bar. The rings still exist.  

\section{Summary} 
\label{sec:Summary}

We simulated the morphological features that result from two consecutive pairs of CCSN jets, in which the narrower jets of the second pair catch up and interact with the structure formed by the first pair of wide jets. We launched the jets from the inner volume of an envelope-stripped massive star collapsing in the framework of the JJEM. According to JJEM (Section \ref{sec:intro}), the newly born NS launches several to about twenty pairs of jets that explode the star. Because of severe numerical limitations, we simplified the present simulations as follows (Section \ref{sec:Numerics}). (1) We launch the jets at thousands of km from the center (compared with the launching zone of the jets of tens of km). (2) We did not include the gravity of the newly born NS and the gas. To justify this, we set up very energetic jets that accelerate the gas to well above the escape velocity from the interaction zones between the jets and the core material. (3) We launch only two pairs of jets. Additional pairs of jets, between the first and second \citep{AkashiSoker2026}, or later,  will expel material that the two pairs we simulated did not. 
For these simplifications, we examine the structure of the outer regions only in the emission integral maps, removing the inner dense gas that we did not explode, as we did not launch other jets (left for later studies). Despite these simplifications due to numerical limitations, our simulations reveal some morphological features that such pairs of jets might form. The most prominent is a pair of opposite circum-jet rings (Figures \ref{fig:em_angles_A} - \ref{fig:em_run_M}).

We present the evolution of the jet-jet interaction in Section \ref{sec:Results}, presenting 2D density (Figures \ref{fig:density_evolution} and \ref{fig:density_evolution_2}), three-dimensional view (Figure \ref{fig:3D}), and velocity (Figure \ref{fig:velocity_map}) maps, and the emission integral (according to equation \ref{eq:EI}; Figures \ref{fig:em_angles_A} - \ref{fig:em_run_M}). We find that circum-jet rings form commonly and require no fine-tuning for this set of jet pairs. This is compatible with observations that reveal rings in about a third of CCSNRs with well-resolved images (Section \ref{sec:Rings}).  

At high inclinations, i.e., $\simeq 70^\circ -90^\circ$ of the jets' axis to the line of sight (lower panels of Figures \ref{fig:em_angles_A} - \ref{fig:em_run_M}), the projection of each ring on the plane of the sky forms two bright zones, where the rings cross the plane of the sky. There are fainter filaments somewhat closer to the center of the SNR than the bright zones, and between them. This general structure that we simulated qualitatively reproduces the `nose' of SNR G46.8-0.3 that we define in the upper panel of Figure \ref{Fig:SNRG46FigureRadio}, as we present in Figure \ref{Fig:SNRG46FigureCompare}. We find that jet shaping explains the morphology of SNR G46.8-0.3 much better than interaction with molecular clouds; in the lower panel of Figure \ref{Fig:SNRG46FigureRadio}, we list the problems with the ambient-cloud model in accounting for SNR G46.8-0.3's morphology.
   
At lower inclinations (upper panels of Figures \ref{fig:em_angles_A} - \ref{fig:em_run_M}), the rings appear very clearly, exhibiting both the bright ring and the inner faint zone. We find some similarities between our simulations and the CCSNR G11.2-0.3, which has two prominent rings on its outer shell. Due to the numerical limitations that we described above and in Section \ref{sec:Numerics}, we can compare only structures on the outer shell. Figure \ref{Fig:SNRG46FigureG11Compare} shows these similarities.
 We recall that the emission integral maps mimic well the X-ray relative intensity, but not accurately the radio maps (Section \ref{seubsec:EI}). Therefore, in those cases, the comparison is more on the qualitative side. 

 We note that jets in the framework of the magnetorotaional explosion process, jets from a black hole in collapsar, and jets launched long after the explosion, as \cite{Gasealahweetal2025} suggested for Circinus X-1, can also shape pairs of rings. However, these processes do not account for the point-symmetric morphology of some CCSNRs. In \cite{SokerAkashi2025}, we argued for two or more misaligned pairs of jets that shaped Circinus X-1, and from that argued for the JJEM. Here we note that SNR G11.2-0.3 has two misaligned symmetry axes \citep{Soker2025G11}, i.e., it has a point-symmetric structure. Therefore, we cannot argue that in all cases rings in CCSNRs are due to the JJEM.  

We consider the agreement between our simulations of energetic jet pairs participating in the explosion process and the observations of these two CCSNRs,  particularly in the point-symmetric SNR G11.2-0.3,   as adding some support for the claim that the JJEM is the primary explosion mechanism of CCSNe.

\section*{Acknowledgements}

 We thank an anonymous referee for very detailed comments that brought us to clarify many points and processes in our study. A grant from the Pazy Foundation 2026 supported this research. 
NS thanks the Charles Wolfson Academic Chair at the Technion for the support.

\newpage





 \bibliography{BibReference}{}
  \bibliographystyle{aasjournal}

\end{document}